\documentclass{aa}
\usepackage{graphicx}
\usepackage{txfonts}
\usepackage{deluxetable}
\usepackage{lscape}

\usepackage{epsfig}
\usepackage{textcomp}

\begin{document}

\title{Unveiling the oldest and most massive galaxies at very high redshift}

\author{G. Rodighiero\inst{1},
       A. Cimatti\inst{2}, A. Franceschini\inst{1}, M. Brusa\inst{3},
       J. Fritz\inst{1} and M. Bolzonella\inst{4}}
   \offprints{G. Rodighiero}

   \institute{
   Dipartimento di Astronomia, Universit\`a di Padova, Vicolo Osservatorio 2, I-35122, Padova, Italy
         \email{giulia.rodighiero@unipd.it}
         \and
   Dipartimento di Astronomia, Universit\`a di Bologna, Via Ranzani 1, I-40127, Bologna, Italy
         \and
   Max Planck Institut f\"ur Extraterrestrische Physik, Postfach 1312,
   D-85741 Garching bei M\"unchen, Germany
         \and
   INAF-Bologna, Via Ranzani, I-40127 Bologna, Italy}

   \date{Received date; accepted date}

%\maketitle

\titlerunning{High redshift massive galaxies}

\abstract{%Context
The identification and characterization of massive galaxies over a wide
redshift range allow us to place stringent constraints on the cosmic
history of galaxy mass assembly and on current models of galaxy formation and evolution.
%}
%{%Aims
This work explores the existence of high redshift massive galaxies 
unveiled with {\em Spitzer}+IRAC, but missed by conventional selection techniques 
based on optical and near-infrared observations. To this end, we
use multi-wavelength imaging data available for the GOODS-South
field (130 arcmin$^2$), and select a flux-limited sample from the
IRAC 3.6 $\mu$m image to $S_{3.6}\geq 1.8 \mu$Jy (m(AB)$<$23.26).
In order to identify the most extreme objects and to complement previously 
published selections
in this field, we confine our study to the galaxies undetected by the
optical HST+ACS imaging and close to the detection limit of the
$K$-band image ($K>23.5$ AB). Our selection unveiled 20 galaxies
on which we performed a detailed analysis.
%}
%{%methods
For each galaxy, we built a Spectral Energy Distribution (SED) based on
optical-to-8$\mu$m photometry. The SEDs were then used to estimate the
photometric redshifts and to derive the main galaxies physical
properties.
% of these through a $\chi^2$ fitting procedure.
Further constraints were also obtained from the available X-ray and 24$\mu$m data.
%}
%{%Results
The majority of the sample (14 out of 20) sources show degenerate/bimodal solutions for
the photometric redshifts. These can either be heavily dust-enshrouded
($A_V\sim2-4$) starbursts at  $2<z<3$ with bolometric luminosities
$L_{IR}>10^{12} L_{\odot}$, or massive post-starburst
galaxies in the redshift interval $4<z<9$ with stellar masses of $\sim 10^{11}M_{\odot}$.
The remaining six galaxies present a less ambiguous
photometric redshift: with the exception of one low-$z$
dusty source, these latter objects favour a low-extinction solution, with
four of them showing best-fit photo-z solutions at $z\sim4$. One galaxy, ID-6, the only 
source in
our sample with both an X--ray and a 24 $\mu$m detection, might be an extremely
massive object at $z\sim 8$ detected during a post-starburst phase with concomitant 
QSO activity responsible for the 24 $\mu$m and X-ray emissions 
(although a lower-$z$ solution is not excluded).
%}
%{%Conclusions
Our investigation of \textit{Spitzer}-selected galaxies with very red
SEDs and completely undetected in the optical reveals a potential
population of massive galaxies at $z\geq4$ which appear to include significant
AGN emissions.
These sources may be the oldest stellar systems at $z\sim4$, given that the estimated ages are close to the age of the Universe at that redshift.
We found that these, previously unrecognized, optically obscured objects
might provide an important contribution to the massive-end
($M>10^{11}M_{\odot}$) of the high-$z$ stellar mass function and they
would almost double it.
Our suggested evidence in these mature high-$z$ galaxies of the
widespread presence of hidden AGNs may have important implications for
galaxy formation, due to their feedback effects on the surrounding ISM.
}
%\keywords {galaxies: high-redshift -- galaxies: formation -- cosmology:
%  observations -- infrared: galaxies }

\maketitle

\section{INTRODUCTION}
\label{intro}

Early attempts to estimate the luminosity and mass functions of galaxies have
revealed a surprisingly low rate of evolution of the stellar mass between the
present epoch and redshifts $\sim$1, and particularly so for the most massive
systems (Fontana et al. 2004; Bundy et al. 2005; Franceschini et al. 2006).
Deep extensive near-IR surveys have recently confirmed the presence of
a numerous population of already massive galaxies at $z\sim 2-4$ (e.g.,
Franx et al. 2003; van Dokkum et al. 2003; Cimatti et al. 2004; Daddi et al.
2004; Yan et al. 2004; Le Fevre et al. 2005; Papovich et al. 2006).

However, very little is currently known about the existence, number density
and properties of massive galaxies beyond $z\sim4$. Mobasher et al.
(2005) have recently reported the discovery of a very massive ($M>10^{11}\
M_\odot$) galaxy in the GOODS-South field at $z\simeq 6.5$, although a
lower-$z$ solution has been suggested by Dunlop et al. (2006). Using the
Early Data Release (EDR) by the Ultra Deep Survey (UDS) component of the
UKIRT Infrared Deep Sky Survey (UKIDSS), McLure et al. (2006) have found
nine Lyman-break galaxy candidates at $z\sim 5-6$ with stellar masses
larger than $5\times 10^{10}M_\odot$. The formation epochs of such massive
galaxies, and of their stellar population contents, cover a
wide range of redshifts, whose upper boundary extends well into the
re-ionization epoch (Panagia et al. 2005).

\begin{figure*} %[ht!]
\centering
\psfig{file=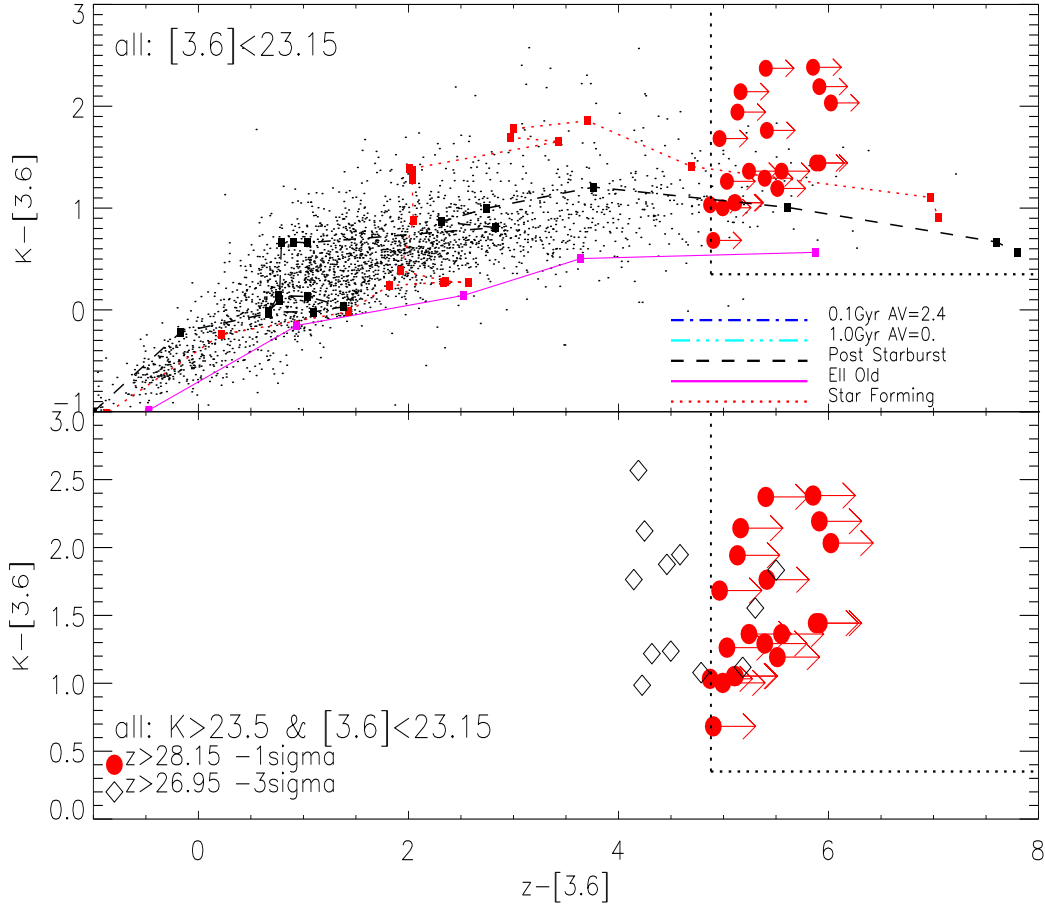,angle=0,width=15cm,height=13cm}
\caption{
Plots of the $K-[3.6]$ versus $z-[3.6]$ colours for our reference GOODS/CDFS
IRAC-selected sample with $S_{3.6}>1.8\ \mu$Jy (small black dots in the top
panel). The objects from our high-redshift sub-sample correspond to the big
red circles with $z$-band limits.
Data are compared with the colours of five spectral templates:
a 10 Gyr old and passive elliptical (solid line), a star-forming
(dotted line) and a post-starburst galaxy (dashed),
a single Simple Stellar Population (SSP) with an age of 1 Gyr
(unextinguished, three-dots-dashed line) and a younger extinguished SSP
of 100 Myr ($A_V=2.4$, dot-dashed line).
The predicted colours are shown as a function of redshift starting
from $z=0$ with increasing steps of 0.5 (the $z=0$ values fall typically in
the bottom-left part of the panel).
The bottom panel zooms onto the colours of our candidate very-high-redshift
objects undetected in the $z$-band down to $z=28.15$ (1$\sigma$ limit),
and an additional sample of objects fainter than $z=26.95$
(3$\sigma$ limit).
}
\label{colcol}
\end{figure*}

A complementary information is offered by very high-redshift quasars
(Fan et al. 2003; Fan et al. 2004) located in highly metal-enriched
interstellar environments (e.g. Dietrich et al. 2003; Freudling et al.
2003). These cases are a clear indication of already advanced evolutionary
stages of the host galaxies at such high redshifts (e.g. Maiolino et al. 2006
and references therein), particularly considering that the identified
atomic species are subject to delayed supernova enrichment, requiring
activity at very high redshifts. Further information comes also from
the detection of thermal dust continuum and molecular emissions in
these quasars including the farthest known QSO at $z=6.42$, SDSS J1148+5152,
indicating the presence of large amounts of dust ($M>10^8\ M_\odot$)
and molecular gas ($M>10^{10}\ M_\odot$) in their circumnuclear environments
(Omont et al. 1996; Omont et al. 2003; Bertoldi et al. 2003; Robson et
al. 2004). The formation of dust grains requires condensation processes in
addition to the stellar activity required to produce the basic elements
(C, Si), and in normal conditions occur in the expanding envelopes of
evolved stars (AGB, late giants, with evolutionary timescales of $>10^8\ yrs$).
If the dust heating came from starburst activity, as it has often been
suggested, that would imply an enormous star formation rate (SFR) of several
hundred to several thousand solar masses per year in the objects.

\begin{figure*}
\centerline{
\psfig{file=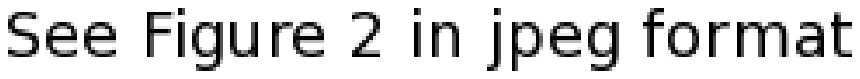,angle=0,height=12cm}
}
\caption{Multiwavelength identification of the twenty 3.6
$\mu$m--selected objects. For each source we report a postage of 5''x5''
in the ACS $z$-band, of 8''x8'' in the ISAAC and IRAC bands and of
17''x17'' in the 24 $\mu$m MIPS band. North is up, East is at left.
}
\label{postages}
\end{figure*}

\section{OBSERVATIONS AND SAMPLE SELECTION}
\label{data}

Finally, the nuclear black-hole mass estimates of the highest-$z$ SDSS quasars
range from several times $10^8$ to several times $10^9\ M_\odot$
(e.g. Fan 2006), such that even with continuous Eddington-limited
accretion they should have started to form at $z > 10$. Although direct
dynamical estimates may suggest a kind of break of the relation
between the black hole and the host galaxy mass at the highest redshifts
(Walter et al. 2004; Peng et al. 2006), it is hard to
imagine that such supermassive black holes do not reside in some kind of
forming massive galactic bulge.

\begin{figure*}
\centerline{
\psfig{file=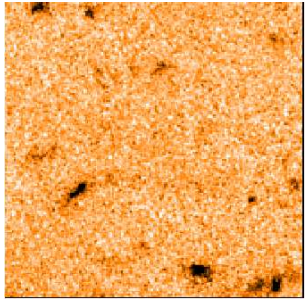,angle=0,width=4cm} }
\caption{
Postage stamps showing the result of the stacking procedure of all ACS
bands ($B$, $V$, $i$, $z$) at the positions of the twenty near-IR dark sources.
No optical signal is detected. The postage has a size of $5''\times5''$.
}
\label{stack}
\end{figure*}

In conclusion, both directly detected high-$z$ galaxies and studies of
the metal enriched circumnuclear media in quasars imply enhanced
star formation activity at very high redshifts ($z>6$) in some specific cosmic
sites. Establishing how frequent the latter are in representative cosmic
volumes would set important constraints on galaxy formation models,
particularly considering the fast decline of the dark matter halo mass
function with increasing redshifts. Indeed, the number densities and
mass functions of host galaxies at the highest redshifts are far from
settled at the moment, while limited information is available only on
the most luminous quasars (e.g. from SDSS \& 2DF).

\begin{figure*}
%[!ht]
\psfig{file=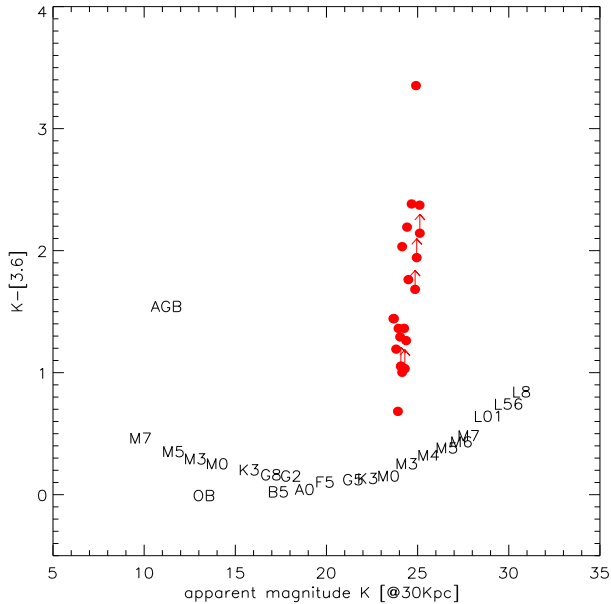,width=9cm,height=8.5cm}
\caption{$K-[3.6]$ colour versus $K$ magnitude plot for our candidate
high-redshift
galaxies (filled circles) and for galactic stars of different spectral types,
as indicated. $K$ magnitudes are calculated by setting stars at a 30 kpc
distance.}
\label{stars}
\end{figure*}

\begin{figure*}
%[!ht]
\psfig{file=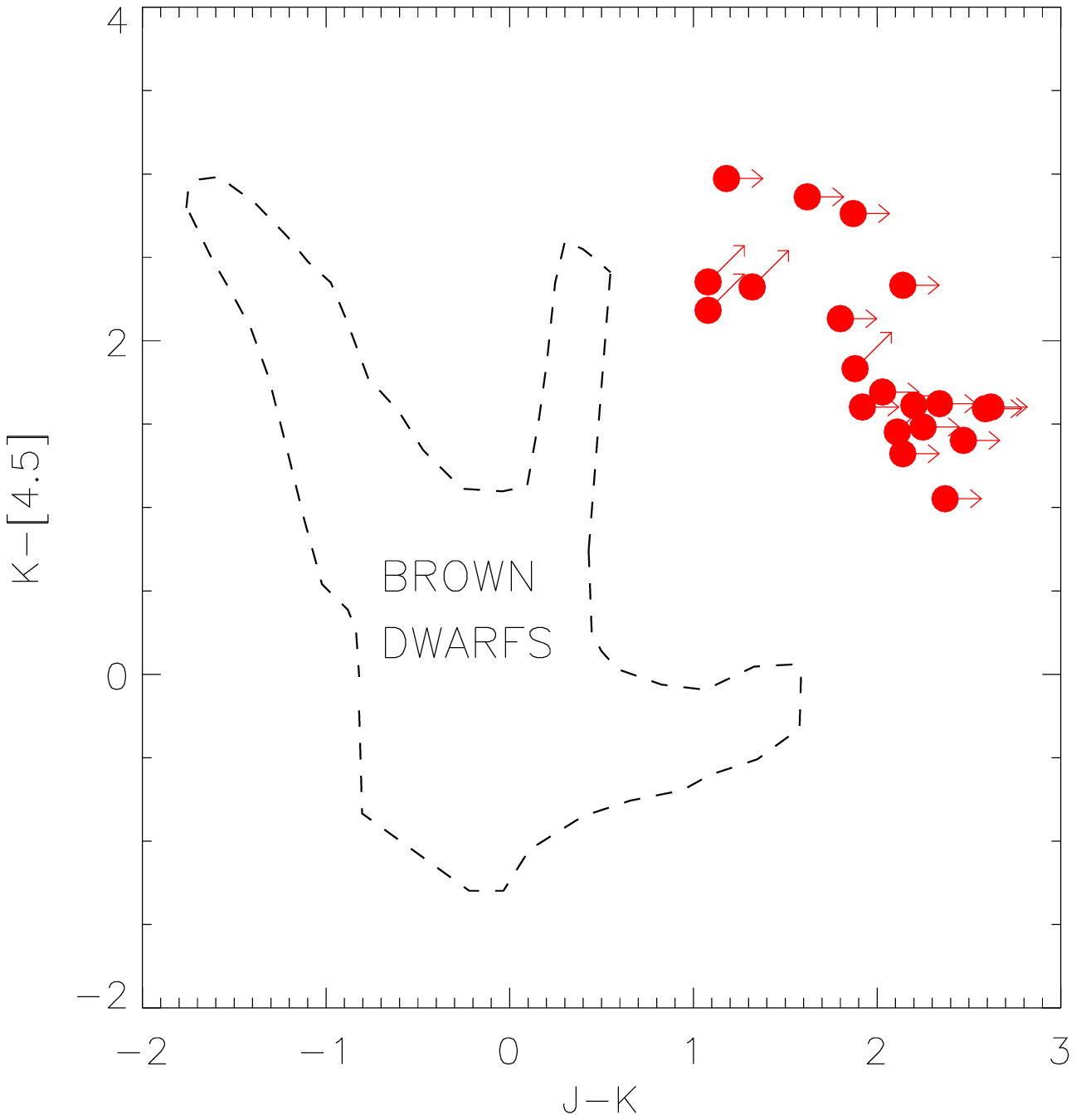,width=9cm,height=8.5cm}
\caption{$J-K$ colour versus $K-[4.5]$ colour for our candidate high-redshift
galaxies (filled circles). The $J$-band magnitudes adopted in this
plot correspond to the 2$\sigma$ value ($J=26.29$).
The dashed region indicates the expected colours for
Galactic brown dwarfs, as recently suggested by Mannucci et al. (2006).
}
\label{b_dwarfs}
\end{figure*}

High redshift galaxies are currently selected using a variety of techniques.
A very successful approach is based on the detection of the spectral
break in the UV continuum blueward of the Ly$\alpha$ due to the intervening
Ly$\alpha$-forest. By construction, this technique is biased towards
star-forming galaxies with rest-frame UV fluxes not strongly reddened
by dust extinction (e.g. Steidel et al., 2003; Bouwens et al. 2006).
This selection technique is biased against selecting high-redshift
galaxies with red spectra, either due to dust extinction or to the
presence of mature stellar populations. Thus, other approaches have
been recently used to select high-$z$ galaxies in a way less affected
by biases. A successful example is represented by near-IR surveys
with or without additional color or photometric redshift selections
(e.g. Cimatti et al. 2002; Franx et al. 2003;  Abraham et al.  2004;
Daddi et al. 2004), or pure flux-limited optically-selected samples
with no color cuts (e.g. Le Fevre et al. 2005), or submillimeter/millimeter
selection of dusty galaxies (Smail et al. 2002, Dannerbauer et al. 2004).

The main question is then: are there high redshift galaxies that are
missed by the current  optical and near-IR selection techniques?

The advent of {\em Spitzer Space Telescope} opened a new possibility
in this respect as it allows to select samples in a spectral region
(3--8$\mu$m) that was not accessible from the ground. The selection
of samples at these wavelengths becomes particularly important
for the specific case of high redshift massive galaxies because
the 3--8$\mu$m selection allows to sample the rest-frame near-IR
of the high-$z$ galaxy Spectral Energy Distributions (SEDs) and is therefore sensitive to their
stellar mass rather than to their star formation activity, and
also much less affected by dust extinction effects.

\begin{figure*}[!ht]
\centerline{
\psfig{file=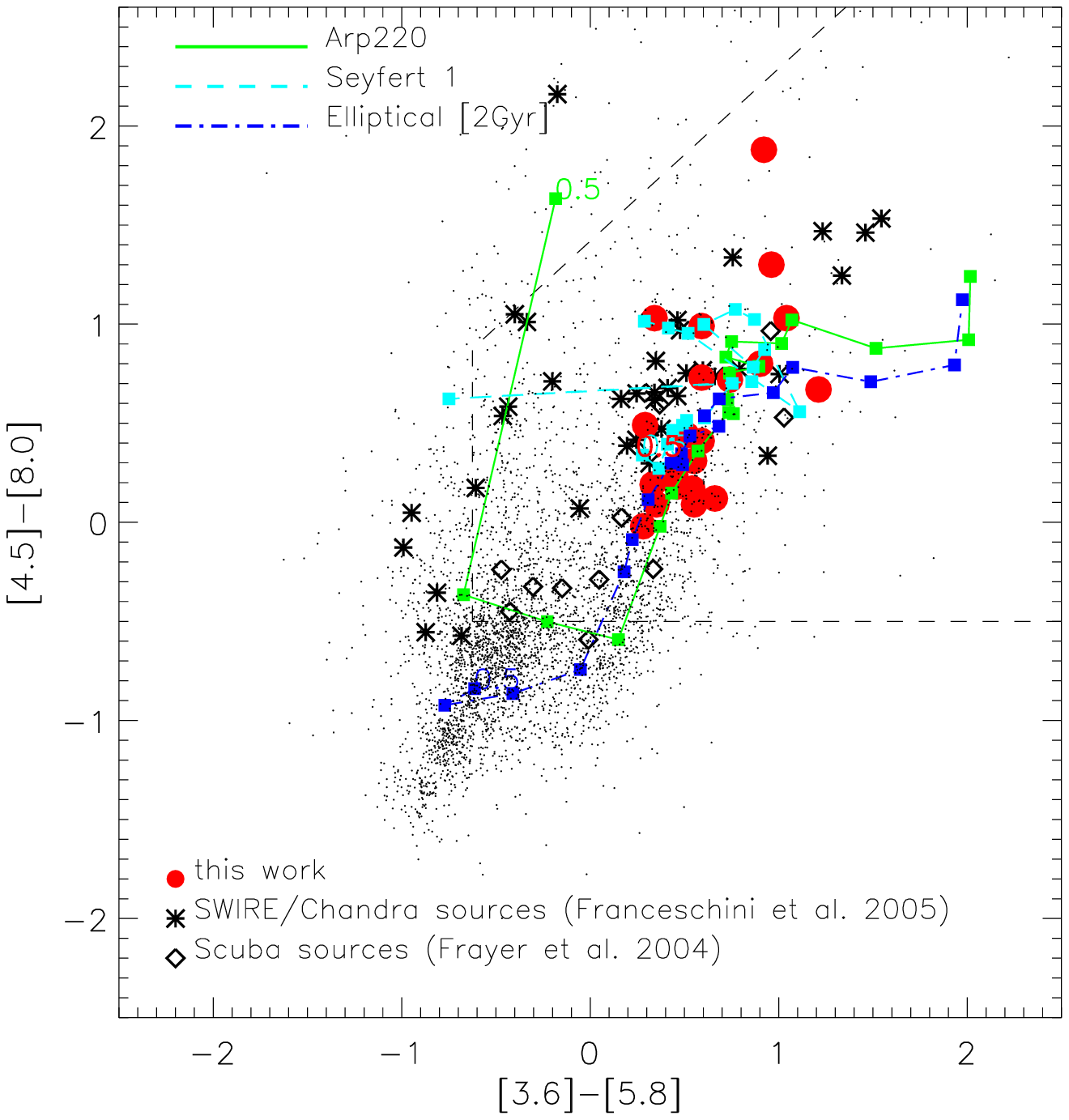,width=12.cm,height=12cm} }
\caption{
The IRAC colour $[8.0]-[4.5]$ is shown against the colour $[5.8]-[3.6]$.
The dashed line marks the region dominated by AGNs in the redshift range
$1<z<3$ according to Lacy et al. (2005).
The near-IR dark objects of our sample are plotted as red filled circles.
For comparison, we report a sample of SCUBA radio-selected sources detected at \textit{Spitzer}
wavelengths (open diamonds, Frayer et al. 2004) and a sample of
Chandra sources detected in the SWIRE survey by Franceschini et al. (2005).
We plot the evolutionary tracks for a reddened starburst (Arp220 -
green solid lines), a Seyfert 1 galaxy (dashed cyan lines)
and a passive elliptical (dot-dashed red lines). The predicted colours
are shown  as a function of redshift with increasing step size of 0.5,
starting from $z=0.5$ (as marked in the plot for each template).
For comparison, we also report the colour distribution (dotted points)
of the GOODS/MUSIC sample (Grazian et al. 2006) with $S(8\mu m)>1\mu$Jy.
}
\label{stern}
\end{figure*}

\begin{figure*}
\centerline{
\psfig{file=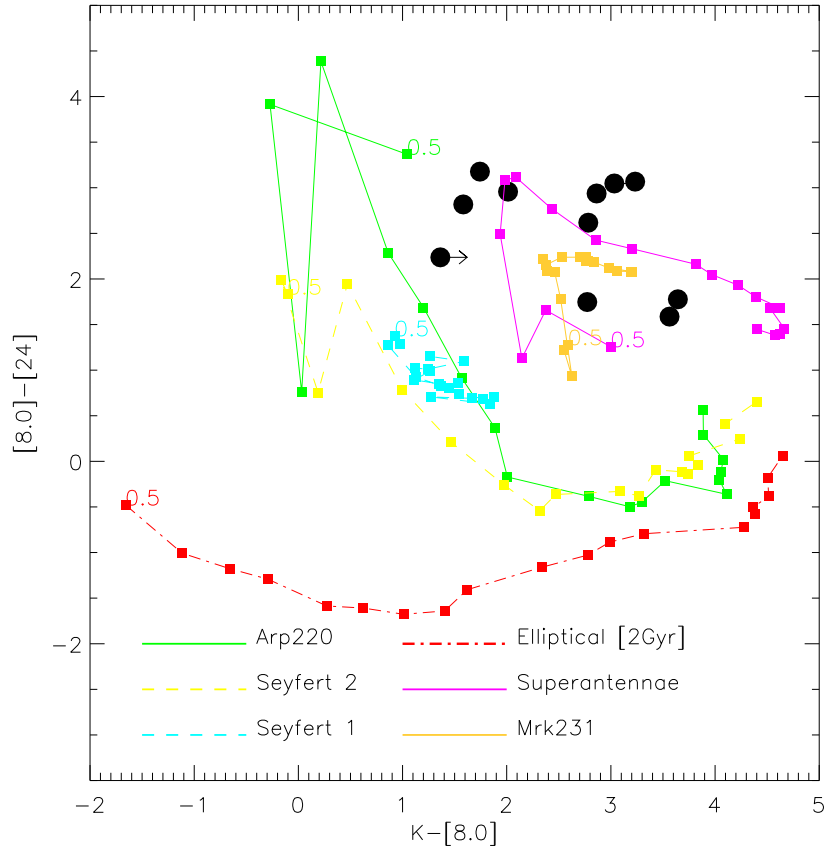,width=12cm}}
\caption{The colour [24]-[8] is shown as a function of the colour
  [8]-[2.2]. Our sources are plotted as big filled circles.       We
  also report evolutionary tracks for starbursts galaxies (Arp220 -
  green solid lines),  AGNs (Seyfert 1 - dashed cyan lines, Seyfert 2
  - dashed yellow line), combined quasar/ultraluminous infrared galaxy
  (ULIRG) sources like the type-1 QSO Mrk 231 (orange solid lines) and
  the type-2 quasar/ULIRG Superantennae South (IRAS 19254s; Berta et
  al. 2003, magenta lines), and a passive elliptical (dot-dashed red
  lines).  The predicted colours are shown as a function
  of redshift with increasing step size of 0.5, starting from $z=0.5$ (as marked in the plot for each template).
}
\label{webb}
\end{figure*}

\begin{figure*}
%[!ht]
\centering
\psfig{file=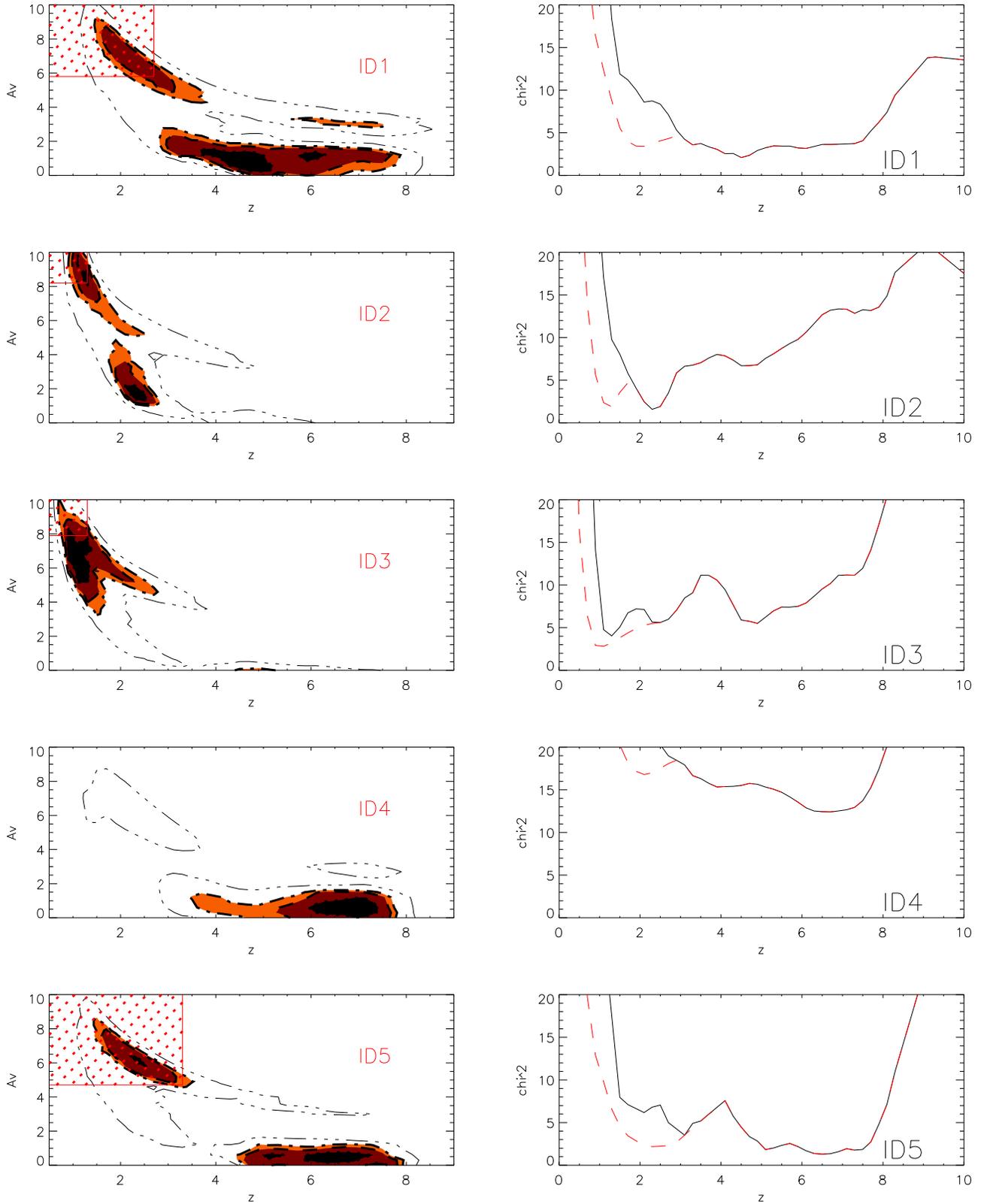,width=18cm}
\caption{Left panel: confidence levels for the photometric redshift 
$z$ and extinction $A_V$ derived from a $\chi^2$ analysis 
are shown models for each source, for the BC03 model.
Dotted, dashed, dot-dashed and three-dot-dashed curves respectively mark the  68\%,
90\%, 95\% and 99.99\% confidence levels
of the $\chi^2$ statistics. For each source, the red shaded area shows regions of the parameter
space which appear to be disfavoured by our analysis in Sect. \ref{z-phot} taking into account the 
24 $\mu$m constraint.
Right panel: the value of the $\chi^2$ is plotted as a function
of redshift. Different curves show the result of using different extinction ranges in the SED
fitting procedure with $Hyperz$ (solid curve: $A_V<6$, red dashed
curve: $A_V<10$).
}
\label{chi_cont1}
\end{figure*}

\begin{figure*}[!ht]
\centering
\psfig{file=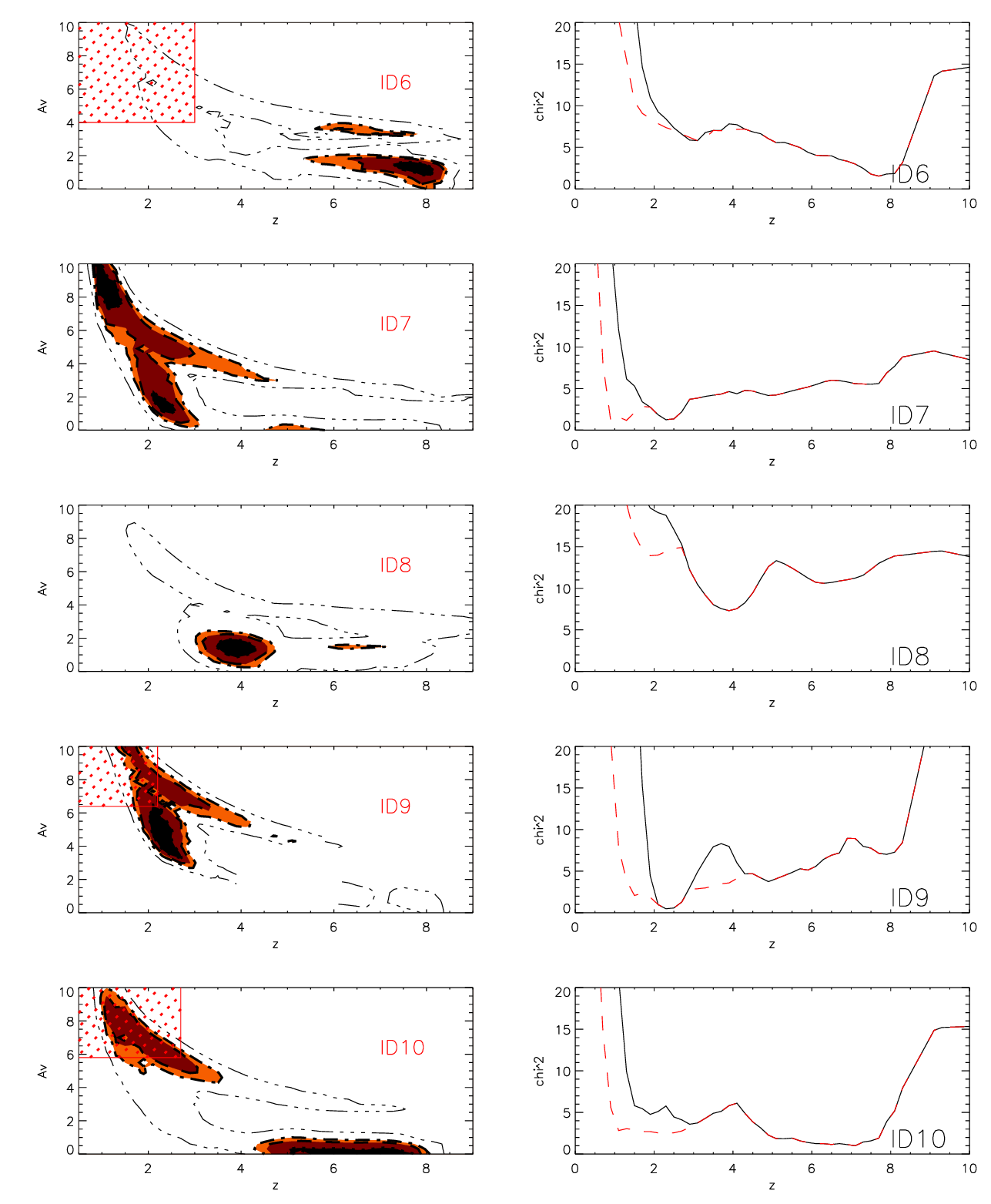,width=18cm}
\begin{center}
Figure~\ref{chi_cont1} (continued)
\end{center}
\end{figure*}

\begin{figure*}[!ht]
\psfig{file=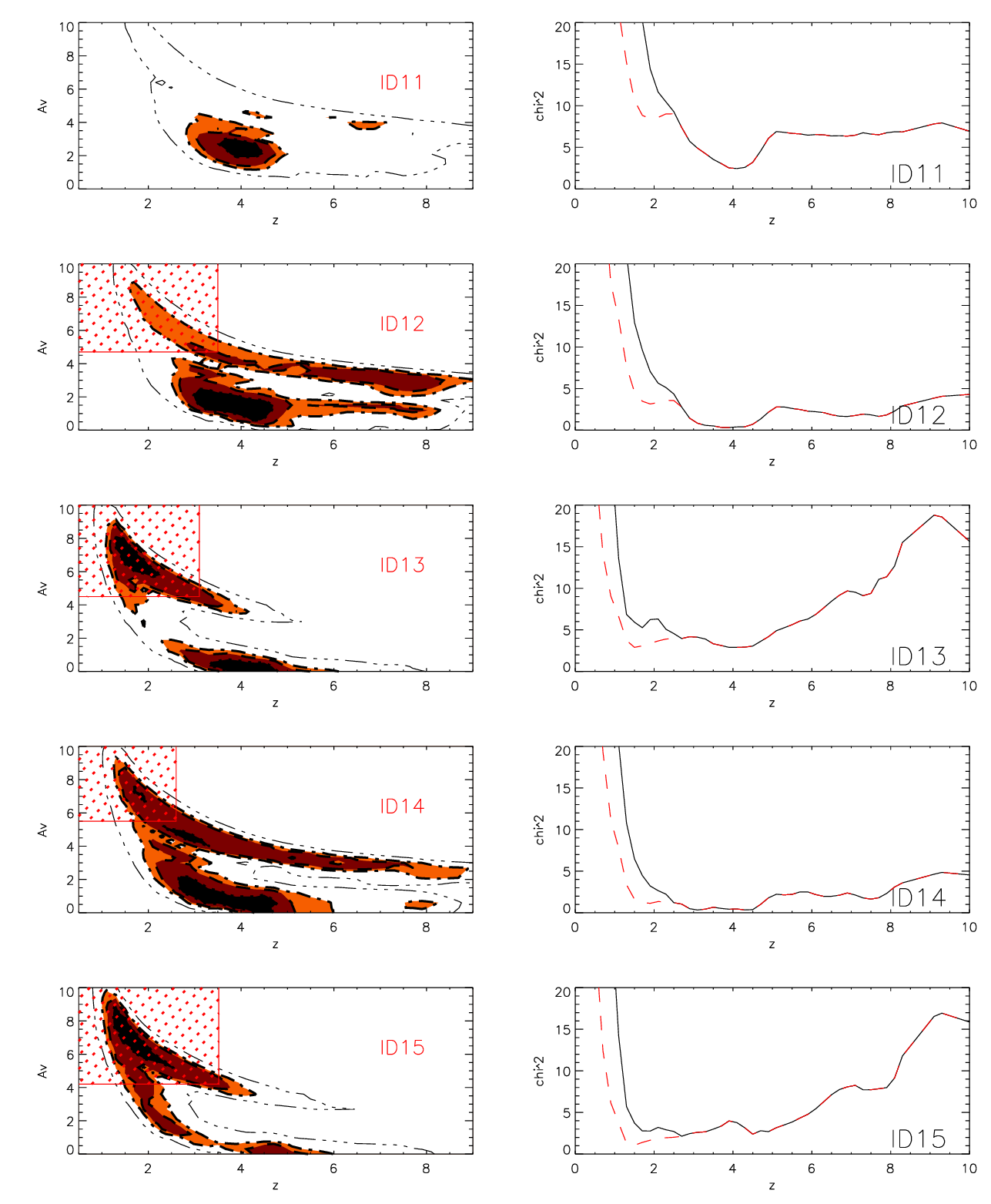,width=18cm}
\begin{center}
Figure~\ref{chi_cont1} (continued)
\end{center}
\end{figure*}

\begin{figure*}[!ht]
\psfig{file=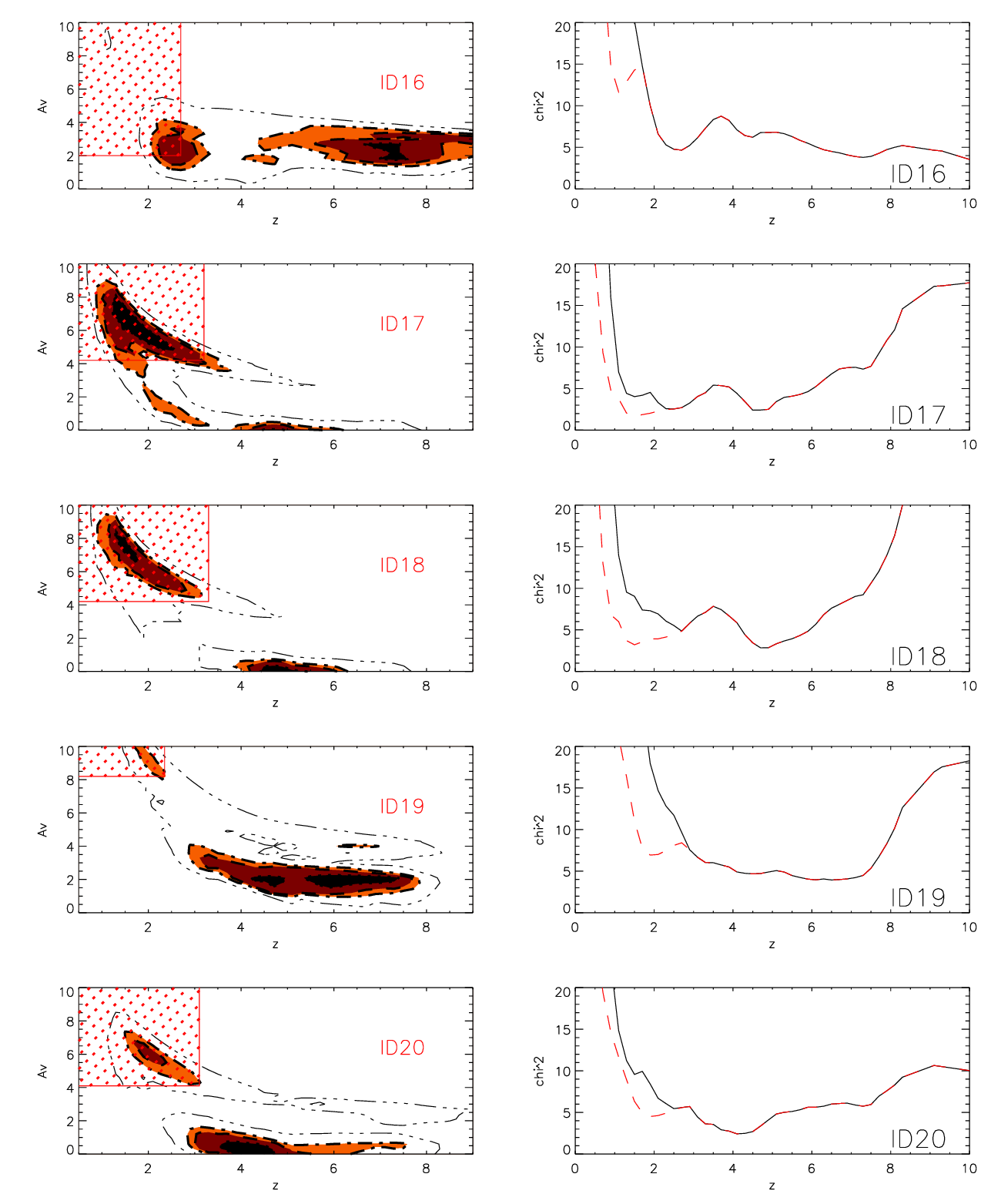,width=18cm}
\begin{center}
Figure~\ref{chi_cont1} (continued)
\end{center}
\end{figure*}

In this work, we explored if there are high redshift massive
galaxies missed by the conventional selection techniques, but
that can be unveiled with {\em Spitzer}. For this purpose, we
used the multi-wavelength deep imaging data available for the
GOODS-South field (130 arcmin$^2$, Giavalisco et al. 2004) and we searched for
extreme galaxies in a way complementary to the other selections
applied in previous surveys in the same field.

The paper is organized as follows. In Section 2 we present the
multi-wavelength photometric data-set and our sample selection criteria.
In Section 3 we discuss the model fitting to the observed SEDs, while
in Section 4 we analyze the infrared colours of our sample.
Section 5 illustrates the X-ray properties of the sample sources. In
Section 6 we report the detection of a candidate massive galaxy at
$z\sim8$. Section 7 presents a general discussion of the statistical
properties of our sample. Finally, in Section 8 we summarize the main
results of this paper. We adopt for the cosmological parameters $\Omega_m$=
0.3 and $\Omega_{\Lambda}=0.7$, and $H_0=70\ Km/s/Mpc$. All magnitudes are
given in the AB system.

With the aim of studying the multiwavelength photometry of extragalactic sources
measured by different instruments, we need to measure the bulk
of the emission from each object in each photometric band.
Only with such kind of approach an SED can be considered reliable.
In our work, we performed aperture photometry in each band.

\subsection{Deep $Spitzer$ Near- and Mid-IR Photometric Imaging}
\label{irac}

As part of the GOODS project, the \textit{Spitzer Space Telescope} has
recently surveyed the CDFS field in the IR between 3.6 and 8.0
$\mu$m using IRAC and in the range 24-160 $\mu$m using MIPS.
The fully reduced data were publicly released by the GOODS team
and are available through the World Wide Web{\footnote
{http://data.spitzer.caltech.edu/popular/goods}}.

In this paper we exploit a galaxy catalogue that we have derived from
the 3.6 $\mu$m IRAC public raw data and recently used to derive the
luminosity and mass function by morphological type (Franceschini et al. 2006).
In order to obtain the most accurate SEDs,
we have recomputed the fluxes of each source by performing aperture photometry
in the four IRAC bands at the positions originally detected in the IRAC
3.6 $\mu$m channel. Assuming that essentially all the sample sources
are seen as point-like by the IRAC $\sim 2\ arcsec$
FWHM PSF imager, we computed with SExtractor (Bertin \& Arnouts, 1996)
the fluxes within a 3.8 arcsec diameter aperture. This choice is
supported by an accurate analysis performed by the SWIRE
team{\footnote {http://data.spitzer.caltech.edu/popular/swire/20050603\_enhanced\_v1/}}.
They constructed color-magnitude diagrams for various types of
objects, in particular main-sequence stars. It was then found that the
scatter in these diagrams is minimized through the use of a
3.8 diameter aperture, and corresponds to roughly twice the
beamwidth. To obtain total fluxes, we then applied the correction
factors indicated by the SWIRE team{\footnote {See Note 2}}.
We have independently verified that the IRAC/SWIRE aperture
corrections are consistent with those derived by fitting the radial
brightness profiles of stars in the GOODS fields.
In the case of extended sources, we used Kron-like magnitudes
(AUTO$\_$MAG output parameter in SExtractor). Our sample turned out to
be $\sim60$\% complete above 1 $\mu$Jy (m$_{3.6}$=23.9), $\sim75$\%
complete above 2 $\mu$Jy (m$_{3.6}$=23.15),
$\sim90$\% at 5 $\mu$Jy (m$_{3.6}$=22.15), and more than $\sim95$\%
above 10 $\mu$Jy (m$_{3.6}$=21.4).

The MIPS public dataset includes calibrated maps and a catalogue of
24 $\mu$m sources with flux densities $S_{24}>80 \mu$Jy. The
photometry is based on a PSF fitting
algorithm, where the SExtractor positions of the IRAC sources are used
as input to the MIPS source extraction process.
The MIPS 24 $\mu$m PSF was generated from isolated sources in the
image, and renormalized based on the aperture corrections published in
the MIPS Data Handbook (v2.1, section 3.7.5, table 3.12).

To extend the 24 $\mu$m sample to fainter fluxes, we have run an independent
PSF fitting algorithm that we already successfully applied in the
GOODS-xFLS/EN1 Science Verification field (Rodighiero et al. 2006).
By these means we have extended the 24 $\mu$m sample down to $S_{24}>
20 \mu$Jy.
% for four additional IRAC sources.

\subsection{Near-IR Ground-based Imaging}
\label{Kimag}

As part of GOODS, near-infrared imaging observations of the CDFS have been
carried out in the $J$, $H$ and $K_s$ bands, using the ISAAC instrument
mounted on the ESO VLT. We made use of the publicly available $J$, $H$
and $K_s$ imaging (version 1.0, released{\footnote {http://www.eso.org/
science/goods/releases/20040430/}} by the ESO/GOODS team in April 2004).
This data release includes 21 fully reduced VLT/ISAAC fields in $J$, $H$ and
$K_s$, covering 130 arcmin$^2$ of the GOODS/CDFS region. It also
includes mosaics of the coadded tiles as single FITS files in $J$
and $K_s$ bands, as well as the corresponding weight maps.
To provide a homogeneous photometric calibration across the entire field,
all images are rescaled to the same zero point (26.0). The final mosaics
have a pixel scale of 0.15".

We measured the $J$, $H$ and $K$ band magnitudes with SExtractor at the
positions of the 3.6 $\mu$m IRAC selected sources through circular apertures
with diameters of 2 arcsec.
Mobasher et al. (2005) found that the photometric curve of growth
converges at this aperture, which represents the
best compromise between convergence of the total flux
and the effects of systematic uncertainty in the background subtraction.

For undetected ISAAC sources, we have initially computed the upper limits to the
flux by measuring the signal in 400 random sky positions.
We used 2-arcsec aperture diameters and calculate the value of the
standard deviation from the distribution of the measured aperture fluxes.
We obtained 1$\sigma$ values of 26.12, 25.82 and 25.12 magnitudes for
the $J$, $H$ and $K$ band, respectively.
However, given that the depth of the ISAAC imaging varies significantly
across the CDFS field, we preferred to compute independent fluxes for
each undetected source. Following the approach of Dunlop et
al.(2006), we perfomed our own manual photometry through a 2-arcsec
diameter aperture at the IRAC-selected positions.

\subsection{ACS/HST Optical Imaging}
\label{acs}

The core of the GOODS project was the acquisition and data reduction
of high-resolution HST/ACS imaging obtained as an HST Treasury Program
(Giavalisco et al. 2004).
The GOODS ACS/HST Treasury Program has surveyed two separate fields
(the CDFS and the Hubble Deep Field North) with four broad-band filters:
F435W(B), F606W(V), F775W(i) and F850LP(z).
In August 2003 the GOODS team released version 1.0 of the reduced, stacked
and mosaiced images for all the data acquired over the five epochs of
observation.
To improve the point spread function (PSF) sampling, the original images,
which had a scale of 0.05 arcsec / pixel, have been drizzled on to images
with a scale of 0.03 arcsec / pixel.

The dataset is complemented by the ACS/HST catalogues released by the
HST/GOODS team in October 2004, containing the photometry in $B$, $V$, $i$
and $z$ bands. The source extraction and the photometric measurements
have been performed by the GOODS team running a modified version of SExtractor
on the version 1.0 images. We have considered
aperture magnitudes using a 1 arcsec diameter. For undetected IRAC sources,
photometry in the ACS $B$, $V$, $i$ and $z$ bands was measured as upper
limits with the same procedure adopted for the ISAAC imaging. We used
1 arcsec diameter apertures and adopted the 3$\sigma$ value as an upper
limit to the flux of undetected sources.

\begin{table*}
\begin{minipage}{200mm}
  \caption{Photometric properties of the sample objects}
 \scriptsize
%\small
\begin{tabular}{ccccccccccccccc}
\hline
\hline
ID & RA & DEC & b & v & i & z & J & H & K & [3.6] & [4.5] & [5.8] & [8.0] & [24]\\
~ & J2000 & J2000 & mag & mag & mag & mag & mag & mag & mag & mag & mag & mag & mag & mag \\
\hline
\hline

1  &  53.12440 &  -27.88268  &  -27.61 & -27.72 & -27.15 & -26.94 & -25.41  & -25.65 &  24.49  & 22.72  &  22.35 &  22.13 & 21.62  &  18.69\\
2  &  53.22620 &  -27.85910  &  -27.61 & -27.72 & -27.15 & -26.94 & -26.19  & -24.97 &  23.95  & 22.58  &  22.32 &  21.92 & 22.20  &  19.03\\
%  5181 &  53.23233 &  -27.85496  &  -27.61 & -27.72 & -27.15 & -26.94 & -24.73  & -24.94 &  24.91  & 21.55  &  21.52 &  21.22 & 21.33  &  18.68\\
3  &  53.06607 &  -27.83178  &  -27.61 & -27.72 & -27.15 & -26.94 & -25.78  & -25.17 &  23.67  & 22.22  &  22.06 &  21.94 & 22.08  &  19.27\\
4  &  53.19752 &  -27.81387  &  -27.61 & -27.72 & -27.15 & -26.94 & -26.86  & -25.06 & -24.86  & 23.17  &  22.85 &  22.83 & 21.82  &  18.78\\
5  &  53.16146 &  -27.81118  &  -27.61 & -27.72 & -27.15 & -26.94 & -25.45  &  24.94 &  24.15  & 22.11  &  21.81 &  21.60 & 21.37  &  19.63\\
6  & 53.15831 &  -27.73355  &  -27.61 & -27.72 & -27.15  &-26.94  &-25.65  &  24.68 &  24.42  & 22.22  &  21.65 &  21.32 & 20.85 &  19.27\\
7  & 53.15552 &  -27.71871  &  -27.61 & -27.72 & -27.15  &-26.94  &-25.37  & -25.06 & -24.30  & 23.26  &  23.02 &  22.71 & 22.93 &  20.70\\
8  &  53.19850 &  -27.92742  &  -27.61  &-27.72  &-27.15  &-26.94  &-26.27  & --*    &  23.92  & 23.23 & 22.86 &  22.64  & 21.87  &   00.00\\
9  &  53.13935 &  -27.89070  &  -27.61  &-27.72  &-27.15  &-26.94  &-26.65  & -25.92 &  25.11  & 22.73 & 22.13 &  21.52  & 21.46  &   19.69\\
10 &  53.14963 &  -27.87682  &  -27.61  &-27.72  &-27.15  &-26.94  &-25.37  & -25.30 & -25.12  & 22.97 & 22.64 &  22.42  & 22.33 &   19.72 \\
11 &  53.14664 &  -27.87103  &  -27.61  &-27.72  &-27.15  &-26.94  &-25.37  & -26.30 &  24.09  & 23.03 & 22.47 &  21.99  & 21.44  &   00.00\\
12 &  53.12938 &  -27.87172  &  -27.61  &-27.72  &-27.15  &-26.94  &-25.39  & -25.20 & -24.95  & 23.00 & 22.57 &  22.26  & 21.85 &   00.00\\
13 &  53.13809 &  -27.86780  &  -27.61  &-27.72  &-27.15  &-26.94  &-26.79  & -26.02 &  24.15  & 23.14 & 22.82 &  22.68  & 22.64  &   00.00\\
14 &  53.20188 &  -27.84421  &  -27.61  &-27.72  &-27.15  &-26.94  &-25.55  & -24.88 &  24.37  & 23.10 & 22.76 &  22.51  & 22.35  &   19.40\\
15 &  53.03842 &  -27.82529  &  -27.61  &-27.72  &-27.15  &-26.94  &-26.05  & -25.08 &  24.26  & 22.89 & 22.56 &  22.35  & 22.39  &   00.00\\
16 &  53.18194 &  -27.81414  &  -27.61  &-27.72  &-27.15  &-26.94  &-26.01  & -24.82 & -24.08  & 23.02 & 22.72 &  22.10  & 20.84 &   17.78\\
17 &  53.12701 &  -27.80455  &  -27.61  &-27.72  &-27.15  &-26.94  &-25.54  & -25.38 &  24.04  & 22.74 & 22.55 &  22.39  & 22.46  &   00.00\\
18 &  53.18047 &  -27.77972  &  -27.61  &-27.72  &-27.15  &-26.94  &-25.54  & -24.91 &  23.70  & 22.25 & 22.10 &  21.80  & 21.86  &   00.00\\
19&  53.11983 &  -27.74309  &  -27.61  &-27.72  &-27.15  &-26.94  &-26.25  & -26.02 &  24.67  & 22.28 & 21.80 &  21.32  & 20.50  &   00.00\\
20&  53.12762 &  -27.70684  &  -27.61  &-27.72  &-27.15  &-26.94  &-25.37  & -24.82 &  23.82  & 22.62 & 22.41 &  22.33  & 21.92  &   00.00 \\

\hline
\hline
\tablenotetext{*}{Source \#8 is out of the $H$ band image.}
\end{tabular}
\end{minipage}
\end{table*}

\subsection{Sample Selection}
\label{selection}

Our main aim is to search for galaxies possibly missed by previous surveys and to find new candidates of high redshift massive systems.
In a recent paper, Dunlop et al. (2006) made use of a $K$-band selected
sample brighter than $K=23.5$ AB mag to investigate the high-$z$ galaxy
population in the GOODS-S field. They conclude that there is no convincing
evidence for any galaxies with $M > 3\ 10^{11} M_{\odot}$ at $z>4$.

In order to search for extreme galaxies ``dark'' at optical wavelengths
and ultrafaint in the near-IR, and to be complementary with the
selection applied by Dunlop et al. (2006) in the same field, we selected
our sample starting from the \textit{Spitzer} IRAC 3.6 $\mu$m image, and applied
the following selection criteria:

(1) flux-limited selection from the IRAC 3.6 $\mu$m image down to
$S_{3.6}\geq 1.8 \mu$Jy (m(AB)$<$23.26).

(2) objects undetected in any of the optical HST+ACS images
at the 1$\sigma$ level (i.e. $z>28.14$)

(3) objects close to the detection limit of the VLT+ISAAC $K$-band deep
image ($K>23.5$ AB, a complementary criterion to that of
Dunlop et al. 2006, including objects with $K<23.5$).

In order to avoid potential photometric uncertainties from the blending
due to the large IRAC PSF, we excluded from our sample a few blended sources.
We found twenty sources satisfying the above conditions. This final sample
was further investigated on ISAAC images: the twenty sources turned out to be
all undetected in the $J$-band (at least at the ISAAC sensitivity), while only
two candidates showed a faint counterpart in the $H$-band image.
Three sources are undetected even in the $K$-band ISAAC image.

To illustrate the effects of our selection criteria, we report in
Figure \ref{colcol} the $K-[3.6]$ versus $z-[3.6]$ colours for our complete
GOODS/CDFS sample with $S_{3.6}>1.8\ \mu$Jy (top panel).
For comparison, we also show here the colour-colour plots of five spectral
template SEDs: a 10-Gyr old passive elliptical, a star-forming and
a post-starburst galaxy\footnote{
The synthetic spectra are taken from the set of template used in Fritz et
al. (2007, submitted). The first is taken as representative of a galaxy during a
post-starburst phase with a second main episode of star formation
--forming $\sim 10\%$ of the total stellar mass-- at $\sim 10^8$ years
while the second is build with a main--impulsive burst at $\sim 13$ Gyr
and a continuous star formation which is truncated at $2\cdot 10^7$ years.},
a solar-metallicity SSP with an age of 1 Gyr (unextinguished) and a younger
extinguished SSP of 100 Myr ($A_V=2.4$).
The predicted colours are shown as a function of redshift starting
from $z=0$ with increasing steps of 0.5 (the $z=0$ value falls in
the bottom left part of the panel).
The bottom panel zooms onto the colours of our candidate very-high-redshift
objects undetected in the $z$-band down to $z=28.15$ (1$\sigma$ limit),
and, for comparison, an additiona sample of objects fainter than $z=26.95$
(3$\sigma$ limit).
The meaning of the symbols is detailed in the caption of Figure \ref{colcol}.

We note that $\sim40$ sources, falling in the colour-colour
region populated by our final sample, escaped our selection criteria
being all brighter in the $K$- and $z$-bands with respect to our imposed
constraints ($K>23.5$ and $z>28.14$).

A summary of the  multiwavelength identifications of the 3.6 $\mu$m objects
is presented in Figure \ref{postages}: for each sources we report here postage stamps
of $5''\times5''$ size in the ACS $z$-band, $8''\times8''$ in the ISAAC and
IRAC bands and of $17''\times17''$ in the 24 $\mu$m MIPS band.

In order to look for any possible faint optical detection, we stacked
together the images of all sample sources in the four ACS bands.
The result of this stacking procedure is presented in Figure \ref{stack},
showing the complete absence of optical signal at the position
of the selected sources.

The photometric data used to construct the spectral energy distributions
discussed in the next Sects. are presented in Table 1. Negative values
correspond to upper limits.
We adopt a common value of 10\% (15\%) of the measured fluxes as
photometric errors for the IRAC (MIPS) bands, in order to reflect
the systematic uncertainties of the instruments
The main contributions to these uncertainties are due to the colour-dependence in the flat field and to
the absolute calibration (see for example Lacy et al., 2005, and the IRAC
and MIPS Data Handbook). For the ISAAC $K$- and $H$-band fluxes,
we adopt as photometric error a standard value of 0.3 magnitudes (as
derived from the standard deviation statistics from our set of Monte
Carlo simulations discussed in Franceschini et al., 2006).

The observational SEDs for all our sample sources are plotted in Figs.
\ref{flz1} and \label{fhz1} below.

\begin{figure*}[!ht]
\centerline{
\psfig{file=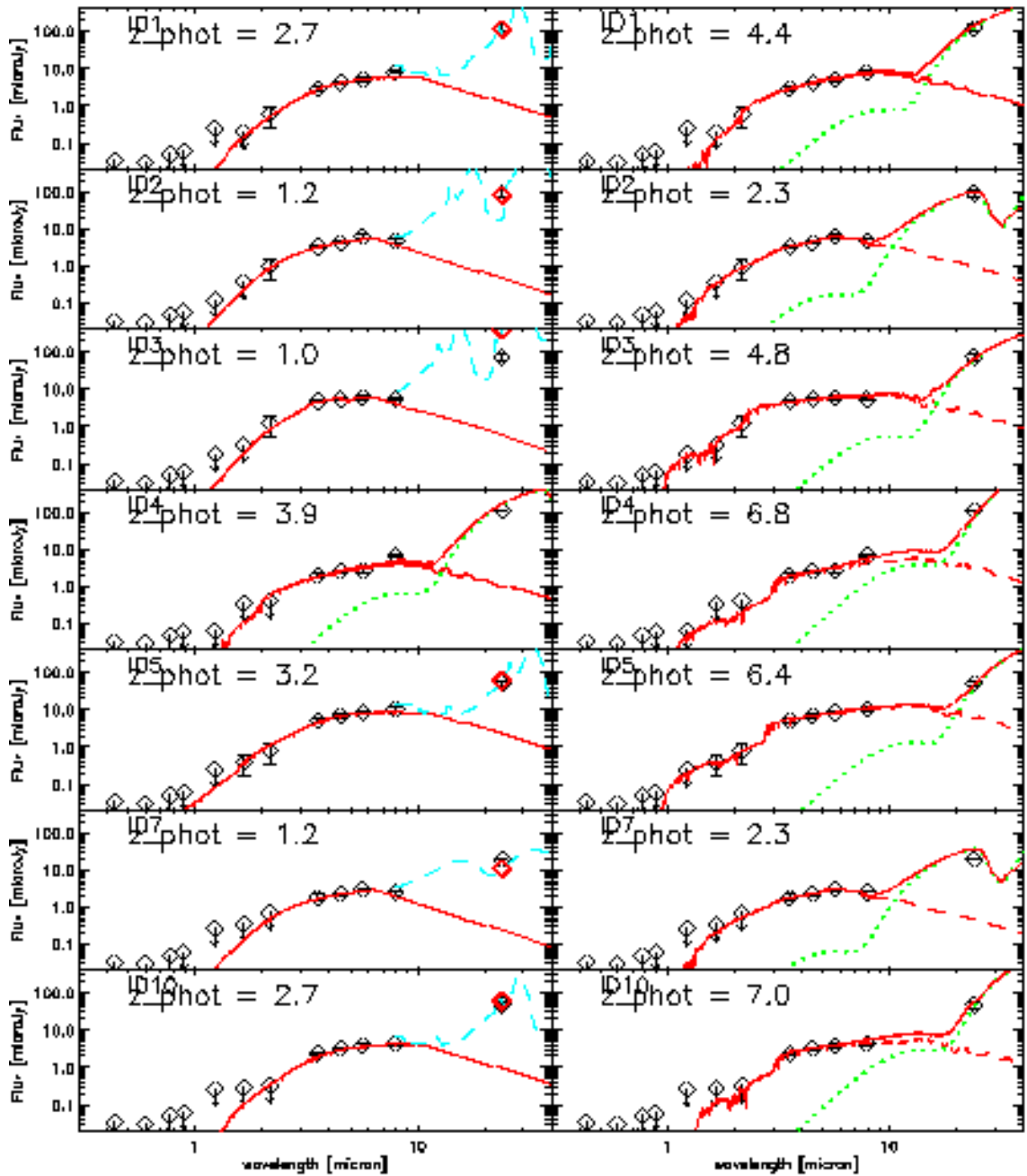,width=16cm}}
\caption{Best-fit models for the sample sources with a bimodal
  solution. In the left panel we report the low-redshift solution.
The observed SED of each source (open diamond)
is shown together with the corresponding low-redshift best-fit solution
(solid red line) indicated by Figure \ref{chi_cont1} and Table 3 for the stellar
  component (up to 8 $\mu$m).
The IR starburst  template that better reproduces the $S(24\mu
  m)/S(8\mu m)$ flux ratio is reported as a dashed cyan line.
The IR spectra have been normalized to match the 24 $\mu$m measurements.
In the right panel we shows the observed SEDs
together with the corresponding best-fit high-$z$ solutions
(dashed red lines, sol. II in Table 3)
In this case, the IR part of the spectra have been reproduced with the spectral
  template of a dusty torus representing the emission of a type-2 AGN
  (green dotted lines, Fritz et al. 2006 model).
The IR spectra have been normalized to match the 24 $\mu$m measurements.
The solid red lines correspond to the sum of the different galaxy components.
}
\label{flz1}
\end{figure*}

\begin{figure*}[!ht]
\centerline{
\psfig{file=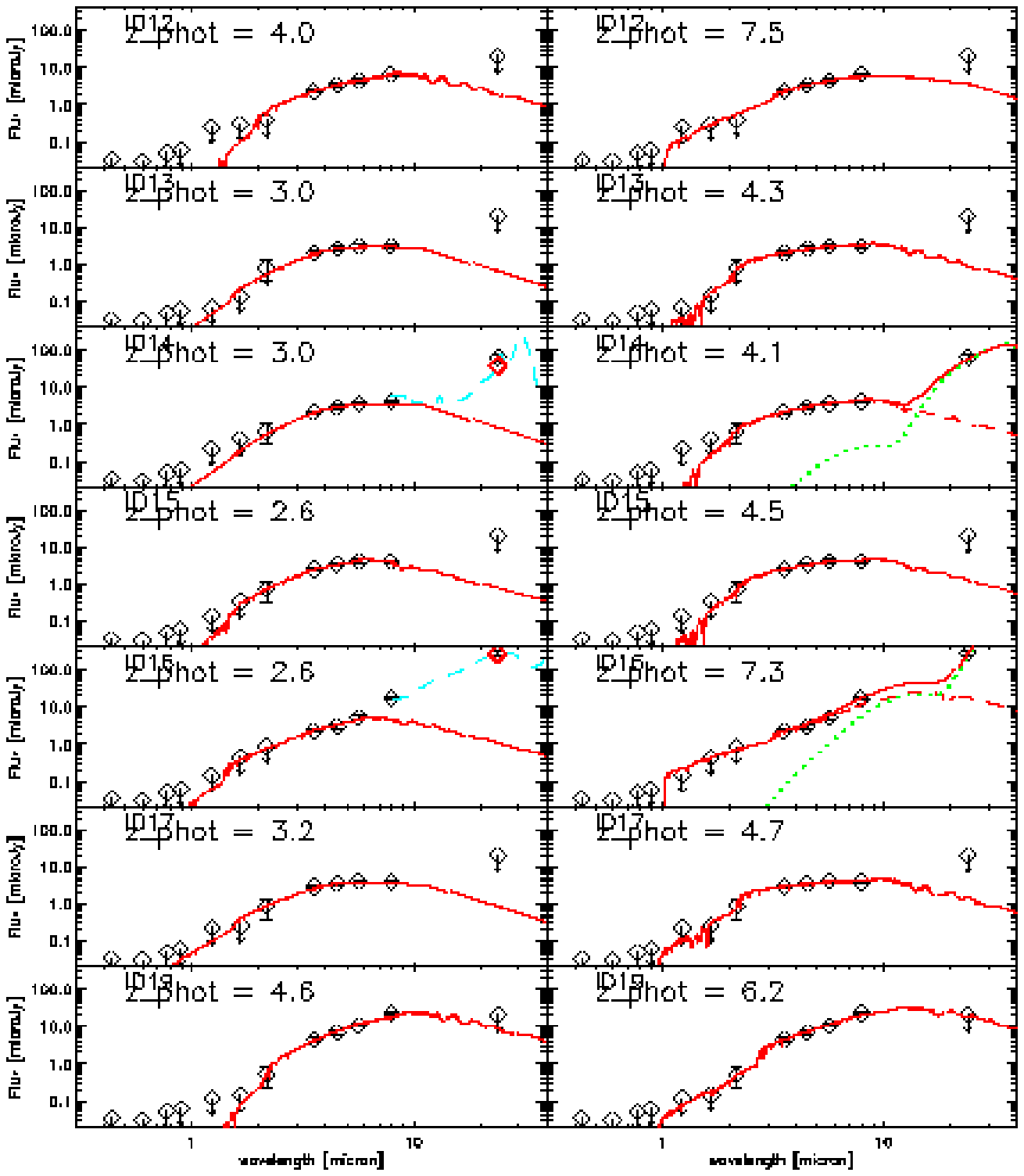,width=16cm}}
\begin{center}
Figure~\ref{flz1} (continued).
\end{center}
\end{figure*}

\begin{figure}[!ht]
\centerline{
\psfig{file=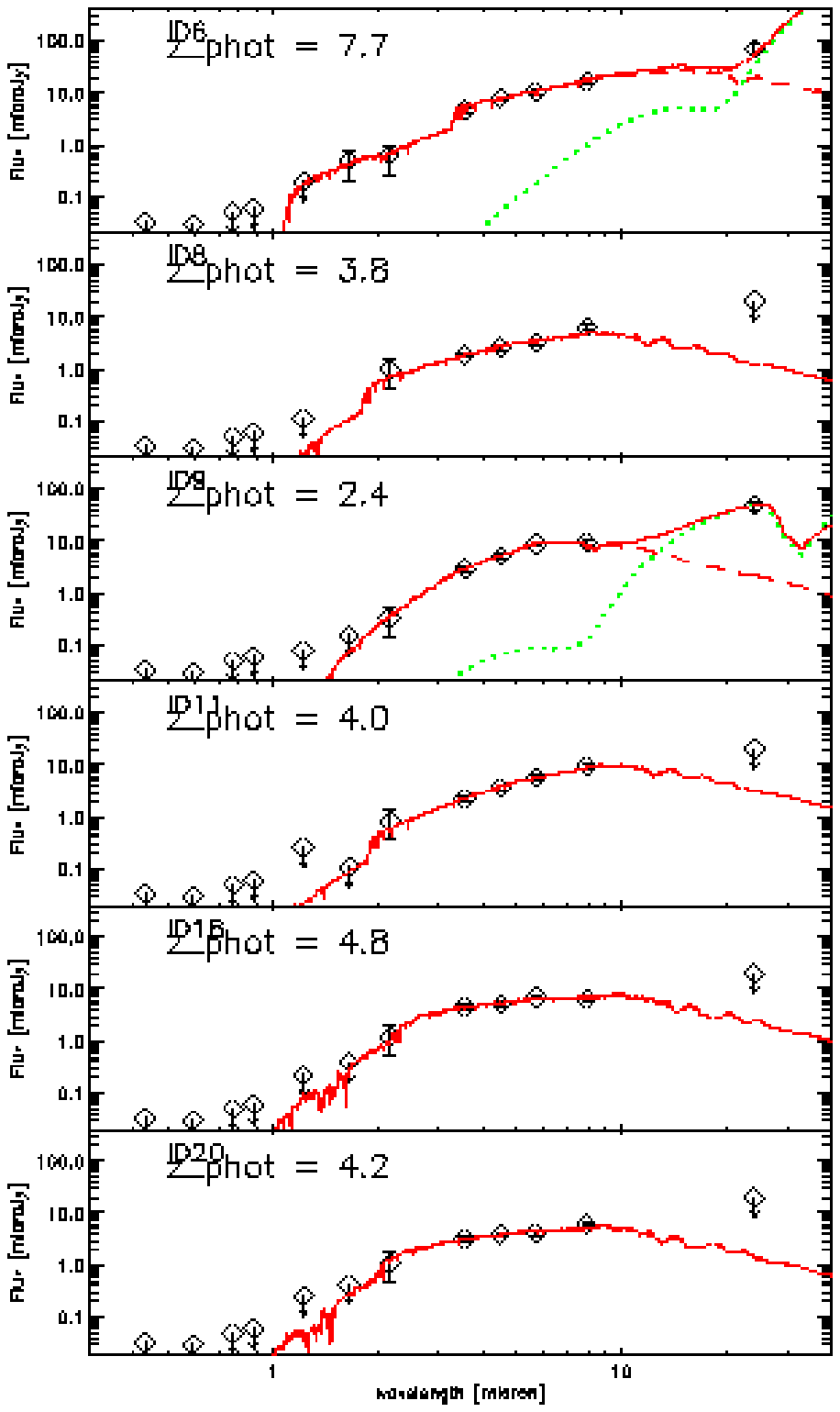,width=9cm,height=18cm}}
\caption{Best-fit models for the sample source with a single
 photometric solution.
 The observed SED of each source (open diamond)
is shown together with the corresponding best-fit solution
(dashed red line) indicated by Figure \ref{chi_cont1} for the stellar
  component (up to 8 $\mu$m).
The IR part of the spectra have been reproduced with the spectral
  template of a dusty torus representing the emission of a type-2 AGN
  (green dotted lines, Fritz et al. 2006 model).
The IR spectra have been normalized to match the 24 $\mu$m measurements.
The solid red lines correspond to the sum of the different galaxy components.
}
\label{fhz1}
\end{figure}

\subsection{Contamination by Galactic Stars}
\label{g_stars}

We have first considered the possibility that some sources within our sample
could be misidentified Galactic cool stars.

We verified that ultra-cool galactic stars (like M and L
dwarfs) show colours (e.g. $K-[3.6]$, see Figure \ref{stars}) that are
in general much bluer than those of our sources. Moreover, the peak of
the stellar emissions of M dwarfs falls shortward
($\lambda \sim 1-2 \mu$m) than observed in the SEDs of our
sample\footnote{We made use of the colour predictions from the stellar model
of Jarrett et al. (1994). We also used the spectral templates of the ISO
standard stars available at http://www.iso.vilspa.esa.es/users/expl\_lib/ISO/wwwcal/.}.

Another possible contaminant is represented by evolved dusty stars, like AGBs
(in particular carbon stars). The spectra of these objects can
in principle reproduce the SEDs of our sample in the 0.4-8.0 $\mu$m
interval range (see again Figure \ref{stars}). However, the detection of several AGB stars in an area of
only 130 armin$^2$ at the magnitude limit of $K<23.5$ is much
in excess of what is e.g. observed in the LMC (Cioni et al., 2006).
In addition, such objects are so bright that they should be moved far away
from our Galaxy (at least $\sim$15 Mpc) in order to match the
observed fluxes of our sample objects.

Young stellar objects (YSOs) may also be considered. Although their
spectra resemble those of our sources, they are usually associated with
extended molecolar clouds in Galactic star-forming regions that are
obviously absent in the GOODS-South field.

An interesting possibility is the potential contamination
by Galactic brown dwarfs. Mannucci et al. (2006) have recently suggested
that a pair of faint $z$-dropout sources ($z-J>0.9$) in the CDFS are compatible
with the expected colours of brown dwarfs. We checked this hypothesis
in Figure \ref{b_dwarfs}, where we report the $J-K$ colour versus the
$K-[4.5]$ colour for our candidate high-redshift galaxies. The $J$-band
magnitudes appearing in this plot correspond to our 2$\sigma$ value
($J=26.29$). The dashed region indicates the expected colours for galactic
brown dwarfs (adapted from Mannucci et al., 2006, that used the stellar
models of Allard et al., 2001). Clearly, all sources in our sample
present $J-K$ colours much redder than expected for brown dwarfs.

\section{INFRARED COLOURS}
\label{colours}

\subsection{Near-IR to IRAC Colours}

To start constraining the nature of our selected galaxies, we have first
compared their mid-IR properties with recently suggested colour-colour diagnostic plots.
While the composite spectra of the stellar populations in
normal galaxies produce SEDs peaking at approximately 1.6 $\mu$m,
the UV to mid-IR continua of AGNs are dominated by power-law emission.
Based on this, Lacy et al. (2004) used IRAC colours from the \textit{Spitzer} First Look Survey to identify AGNs.
A region in the $[4.5]-[8.0]$ versus $[3.6]-[5.8]$ colour plot where AGN are expected to lie is shown
as a dashed-line contour in Figure \ref{stern}.

The twenty near-IR dark objects of our sample are plotted as red filled circles.
For comparison, we also report the colour distribution (dotted points)
of the GOODS/MUSIC sample (Grazian et al. 2006) with $S(8\mu m)>1\mu$Jy.
The colour tracks for a dusty starburst (Arp220 - green solid lines), a Seyfert 1 galaxy (dashed cyan lines)
and a passive elliptical (dot-dashed red lines) are also shown as a
function of redshift, with increasing step size of 0.5, starting from
$z=0.5$ (as marked in the plot for each templates).
All sources fall within the AGN area. However, this is not necessarily
an indication of an AGN dominance in our sample. Indeed,
Fig. \ref{stern} also shows the colour distribution of a sample of
SCUBA radio-selected sources detected at \textit{Spitzer} wavelengths (open
diamonds, Frayer et al. 2004), and the SWIRE/Chandra sample
selected in the EN1 field by Franceschini et al. (2005, asterisks).

Clearly, both starforming galaxies, ellipticals and AGNs at $z>2$ can
reproduce the IRAC colours of our sample, showing a limited diagnostic power by the test.

\subsection{Near-IR -- to -- mid-IR colours}
\label{NMIR}

It is remarkable to note that half of our 3.6 $\mu$m selected sources in our sample (11 out of 20)
reveal a significant 24 $\mu$m excess. Given the MIPS and IRAC
limiting fluxes, this is inconsistent with a
purely passive galaxy at any redshift, and is a clear indication for
some activity taking place in the objects. The nature of this activity
(star formation, AGN or the concomitance of both processes) will be subject of investigation
in the present work.

In Figure \ref{webb} the colour $[8]-[24]$ is shown as a function of
the colour $K-[8]$ (see also Webb et al. 2006), where our sources are
plotted as big filled circles.
We also report here colour tracks for starbursts galaxies (Arp220 -
green solid lines),  AGNs (Seyfert 1 - dashed cyan lines, Seyfert 2 -
dashed yellow line), a combined quasar/ultraluminous infrared galaxy
(ULIRG) sources like the type-1 QSO Mrk 231
(orange solid lines), the type-2 quasar/ULIRG Superantennae (IRAS
19254s; Berta et al. 2003, magenta lines), and a passive elliptical
(dot-dashed red lines). These templates are derived from the spectral
library reported in  Polletta et al. 2006 (see also Franceschini et
al. 2005).
The predicted colours are shown as a function of redshift with
increasing step size of 0.5, starting from $z=0.5$ (as marked in the
plot for each template).

The extremely red infrared colours of our sample are difficult to explain.
The $K-[8]$ colour on the X-axis might be consistent with the SED of a
starburst galaxy, like Arp220, or alternatively with a passively
evolving galaxy, both at $z>3$. Lower-redshift solutions
require additional extinction ($A_V\sim2$) to that of the Arp220
spectrum. The very red $[8]-[24]$ ratio may be consistent alternatively with a
moderate-redshift ($z<3$) dusty starburst (Arp220), or with a dusty
quasar.

The combined set of colours are reproduced only by high
redshift ($z>2.5$) ULIRGs with concomitant QSO activity (the type-2
quasar Superantennae, more marginally by the infrared-luminous type-1 QSO Mrk231).
This colour-colour plot of Fig. \ref{webb} will be considered in later applications.

\section{SPECTRAL ENERGY DISTRIBUTION ANALYSIS}
\label{SEDss}

\subsection{Synthetic Spectral Models and Photometric Redshift Estimates}
\label{z-phot}

We made use of the $Hyperz$ code (Bolzonella et al. 2000) to estimate the
photometric redshifts of each source. The whole broad-band photometric
dataset available to us was exploited, with the exclusion of the 24
$\mu$m flux (the stellar population synthesis model template spectra
do not include dust emission).
The fitting procedure was based on a maximum-likelihood algorithm
and the quality of the fit is investigated by means of a $\chi^2$ statistics.
The code computes the $\chi^2$ for a given number of templates, which
differ for star formation histories, metallicities and ages, and finds the
best-fitting template among them.

We used two stellar population synthesis models to fit the observed SEDs:
Bruzual \& Charlot (Bruzual \& Charlot 2003, hereafter BC03) and the
Maraston (2005, hereafter MA05)  models. In both cases, we assumed
exponentially decreasing star formation rates (SFR) parameterized by a
time-scale $\tau$. We have considered the following  set of values for
$\tau$: 0.1, 0.3, 1, 2, 3, 5, 15, 30 Gyr.
Moreover, we included a case with constant SFR, and a single burst
model corresponding to an individual simple stellar population (SSP).
For the BC03 model we used a Chabrier Initial Mass Function (IMF).
The MA05 evolutionary tracks available to us have been generated with
a Salpeter IMF.
To convert mass estimates obtained with Salpeter IMF to Chabrier IMF we
used a constant value of 0.23 dex (which corresponds to correct
the masses computed with Salpeter IMF downwards by a factor 1.7).
This value has been obtained by comparing the stellar masses estimated with BC03 models built with
different IMFs for an observed spectroscopic sample of galaxies. In this
way we obtained the mass difference expected for galaxies with the same
colours.

The extinction parameter is allowed to span the widest conceivable range of
values $0<A_V<10$ (with a step of $d(A_V)=$0.1).
We assumed the extinction law by Calzetti et al. (2000), and the
metallicity was set to the solar value, in order to minimize the
number of free parameters at play.  We also accounted for Lyman-series
absorption due to HI clouds in the intergalactic medium, following the
prescription of Madau (1995).
Even if the reddening is formally allowed to reach extreme values,
we remind that values of $A_V$ exceeding $\sim$5-6 magnitudes have been observed
only in the inner part (i.e. the central 1-2 kpc) of local luminous IR
galaxies (e.g. Mayya et al. 2004, Poggianti et al. 2001).
In comparison, at high redshift ($2<z<3$) the dusty submillimeter galaxies have
typical average extinction around $A_V\sim2.4$ (Smail et al. 2004, Knudsen
et al. 2005).
Another class potentially including heavily obscured sources is that
of Distant Red Galaxies (DRGs) and dusty Extremely Red Objects (EROs).
However, even in this case the reddening has typical values around
$A_V<\sim$3.0 at $2<z<3$ (Cimatti et al. 2003, Moustakas et al. 2004,
Stern et al. 2006,  Papovich et al. 2006).
An Hyper Extremely Red Object (HERO) has been proposed to lie at
$z$=2.4 with $A_V$=4.5 mag (Im et al. 2002).
However, no convincing evidence for galaxies with a global value of
$A_V$ exceeding 5-6 mag has been reported until know.

\subsection{Additional constraints from 24$\mu$m flux}
With the attempt of reducing the uncertainties in the extinction
estimate, we have considered the MIPS 24 $\mu$m flux in
addition to the 0.4-8 $\mu$m photometric data. The dust emission
spectrum longwards of few $\mu$m (restframe) has been modelled by assuming
the observed IR SED of Arp220, and then compared to the flux
detected  in the MIPS band. The Arp220 IR template corresponds to
the spectrum of an highly absorbed ultra-luminous IR starburst, a
conservative choice for our analysis.

The predicted IR emission for each one of our objects is calculated as the 
difference between the unextinguished and the extinguished optical
emission, assuming that all the flux absorbed by dust is re-processed 
and re-emitted longwards of few $\mu$m. The Arp220 template
is then rescaled in such a way that its bolometric emission between
8 and 1000 $\mu$m coincides with the dust reprocessed luminosity.
Finally, the properly scaled Arp220 SED is K-corrected
and convolved with the MIPS filter response to compare with the
observed 24 $\mu$m flux or flux upper limit (see also Berta et al. 2004 ).
With this procedure, we have been able to use the measured 24 $\mu$m flux as a
constraint on the maximum $A_V$ allowed to each photometric redshift solution.

If compared to other typical LIRG galaxies (e.g. the prototype local
starburst M82), the ULIRG Arp 220 is characterized by a much more peaked 
FIR emission and dust self-absorption.
The choice of such an extreme galaxy as an IR template then implies much
higher bolometric infrared luminosities for a given 24 $\mu$m flux,
providing an indication of the maximun value of
$A_V$ consistent with the observed 24 $\mu$m flux.

The results of the SED fitting analysis are reported in Figure \ref{chi_cont1},
where for each sources we show on the left panel the confidence levels
derived from the $\chi^2$ statistics (averaged over all the available free parameters)
as a function of redshift $z$ and $A_V$, for the BC03 models (the MA05 model
provides equivalent results in terms of photometric redshift solutions, see
Section \ref{mar_bc03}). Dotted, dashed and dot-dashed curves
respectively mark the  68\%, 90\%, 95\% and 99.9\% confidence levels of the 
$\chi^2$ statistics. For each sources, the red shaded area indicates a region of the
parameter space which is disfavored by our comparison with the 24 $\mu$m flux constraint: 
spectral solutions falling in this range, all heavily dust extinguished, tend to produce
24 $\mu$m fluxes in excess of the observed values. Although somewhat model-dependent and
unable to associate a formal significance value, our
analysis is made relatively robust by our reference to the most extinguished object 
- Arp220 -  known at the present cosmic time.
In particular, we have verified that the adoption of other IR spectral templates, like that
of M82 or other non standard starburt galaxies, would imply somewhat
wider extension of such ``disfavoured'' regions (in the sense that even
lower values of the Av parameter would be inconsistent with the
measured 24um flux or the adopted upper limits). 
Note that the squared shape of the shaded area is shown here for 
illustrative purposes only.  

\subsection{Results of the SED fitting}
On the right panel of Fig. \ref{chi_cont1} the value of the $\chi^2$ is plotted as a function
of redshift. Different curves show the result of using different extinction ranges in the SED
fitting procedure with $Hyperz$ (solid curve: $A_V<6$, red dashed curve: $A_V<10$).

As a first result, this analysis indicates the existence of multi-modal solutions
for the majority of our sample sources (14 out of 20, \#1, \#2, \#3, \#4, \#5, \#7, \#10, \#12,
\#13, \#14, \#15, \#16,\#17, \#19), even accounting for the 24 $\mu$m constraint: 
generally these objects have statistically degenerate fits with either
a lower-redshift ($z<4$) and high-extinguished ($A_V\sim3$) SED or with a
higher-redshift ($z>4$) and lower extinction.
Sources \#4 and \#19  have best-fit solutions at $z>\sim4$.

Six sources (\#6,\#8,\#9,\#11,\#18 and \#20) show a single preferential
best-fit solution if we consider the constraint from the MIPS emission
and if we limit the analysis to the 95\% confidence level. This number reduce to three sources
(\#6, \#8 and \#11) if this constraint is not taken into account. With the exception of the
dusty source \#9, all these objects favour a low-extinction solution: four of them lie at $z\sim4$.
Only source \#6 favours an higher redshift best-fit ($z\sim8$, see dedicated discussion in Sect. \ref{z8}).
In any case, it should be stressed that, even in the case of single best-fit SEDs, the 
the $z\sim4$ solutions are only "formally preferred", and alternative lower-z 
fits cannot be ruled-out with any confidence.

Stellar masses are computed using an adapted version of the $Hyperz$ code
which performs SED fitting at a fixed redshift.
We report in Table \ref{tab_highz} a summary of the best-fit parameters
computed by $Hyperz$.
Only the solutions matching the 24 $\mu$m constraint are
taken into account in the following.
The upper panel of the table reports the objects with degenerate bimodal redshift
solution. In this case, for each parameter we present the lower redshift
primary solution (sol. I) and the secondary solution (sol. II) at higher
redshift. Moreover, for each physical parameter
we report a range of values: the value reported on the left side of the
interval corresponds to the BC03 model expectation, while the
value reported on the right side of the interval corresponds to that
of the MA05 model.
In the lower panel of the table we report the same information for the
six sources that we consider to have a single photometric redshift.

For sources with a bimodal behaviour, we compare in Figure
\ref{flz1} their observed SEDs (open diamonds) with the corresponding low-$z$ best-fit solutions
(left panel, solid red lines) reported in Table 3 (sol. I), for the BC03 model.
Similarly, in the right panel the observed SEDs are shown together with the corresponding best-fit high-$z$ solutions (dashed red lines, sol. II in Table 3).
Similarly, in Figure \ref{fhz1} we report the best-fit solution for the six non-degenerate objects (solutions presented in the lower panel of Table 3).
%}

\subsection{Comparison of the Maraston and Bruzual \& Charlot Models}
\label{mar_bc03}

As mentioned in the previous Section, we have considered two independent evolutionary
population synthesis models, by MA05 and BC03, in order to check the stability of our results.
In Table 3 we report the results for both models.
The MA05 model accounts in detail for the contribution of the thermally-pulsating
Asymptotic Giant Branch (TP-AGB) phase of stellar evolution.
The TP-AGB phase is calibrated with local stellar populations and is the dominant source
of bolometric and near-IR energy for stellar populations in the age range 0.2 to 2 Gyr.
The two models have an underlying different treatment of convective overshooting and
Red Giant Branch stars.

In a recent paper, Maraston et al. (2006) have tested their code on a
sample of high-redshift early-type galaxies ($1.4<z<2.5$), for which
the MA05 models indicate younger ages (by a factor up to 6) and lower
stellar masses (by $\sim$60\% on average) with respect to results obtained with the models of BC03.

For our sample we found that the photometric redshifts obtained with the BC03 models are very similar to those obtained with the MA05 model. The bimodal distributions in the ($z,A_V$) parameter
space and the $\chi^2$ distributions are essentialy equivalent for the two independent models. 
A similar situation applies for the remaining sources. 
As can be seen in the second and in the
third columns of Table 3 ($z$-phot: sol. I and sol. II), the scatter in
the photometric redshift estimates with MA05 and BC03 is very small at
both low and high redshifts. When considering the solution at lower
redshift, the mean ratio of the photometric redshifts derived from BC03
and MA05, respectively, is 1.04 with an RMS of 0.12, and for the solution
at higher redshift the mean ratio of the photometric redshifts is 1.07
with an RMS of 0.18.
A slightly larger discrepancy is found for the reddening, with no
clear trends.

The most interesting physical parameters are the age and the stellar mass
predicted by the two models. In the lower redshift solution the MA05 model
predicts older ages than BC03 for only 2 out of our 14 objects.
For the higher-$z$ solutions (including the six non-degenerate
objects), 11 out of the 20 sources with a high-$z$ prediction are older for the MA05.
As for the stellar masses, we confirm the results of
Maraston et al. (2006): in the low-$z$ solutions (mean $z\sim3.2$) the BC03
models predicts in general 50\% higher masses on average, and a similar trend 
is found for the high-$z$ (mean $z\sim5.5$) fits ($\sim60$\%).

\subsection{Nature of the 24$\mu$m Emission}
\label{24mu}

As discussed in Sect. \ref{NMIR}, the excess 24 $\mu$m emission often
observed in the SEDs of our galaxies (Figs. \ref{flz1} and \ref{fhz1})
requires the presence of ongoing star formation or AGN activity.

In the extinguished low-redshift case ($1<z<3$), the rest-frame MIPS band samples
the emissions by warm dust and PAH molecules. Both a starburst and an
AGN emissions are then possible explanations of the 24 $\mu$m flux.
We decided in this case to associate the whole mid-IR flux to a
starburst process and adopted the Arp220 SED template (from the
spectral library by Polletta et al., 2006). As described in
Sect. \ref{NMIR}, this template has been used to compute the IR
luminosity (8-1000 $\mu$) of each object matching the observed 24
$\mu$m flux (dashed cyan lines in Figure \ref{flz1}).

At higher redshifts ($z>3.5$), the low values of the extinction favour a
scenario dominated by dust-free sources. In this case the 24 $\mu$m flux
has been independently modeled with the emission of a dusty torus reproducing
the contribution of a type-2 AGN (Fritz et al. 2006).
Again, in Figures \ref{flz1} and \ref{fhz1}  we show the best-fit solutions at
high-$z$, superimposed with the type-2 spectral templates that better
reproduce the $S(24\mu m)/S(8\mu m)$ flux ratio (green dotted lines).

\begin{figure}[!ht]
\centerline{
\includegraphics[width=0.36\textwidth]{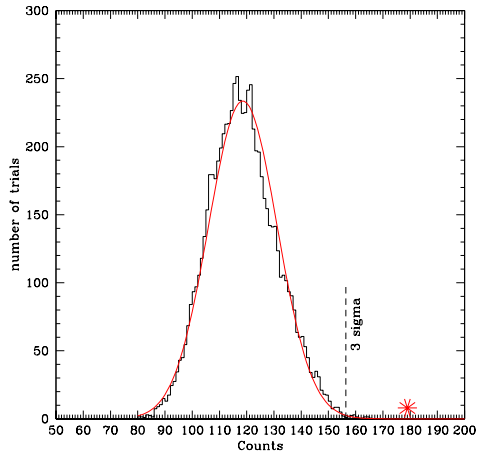} }
\caption{Distribution of background counts in the full (0.5-7 keV) band.
The histogram shows the distribution of background counts
in the source cells derived from the Monte Carlo method
(10,000 trials).
The vertical dashed lines show the 3$\sigma$ fluctuation limit:
stacked detection with a number of counts larger than these values
are considered as ``real'' detections at the 99.9\% confidence level.
The big (red) asterisk represents the counts actually detected from the
sample of 17 individually X-ray undetected sources in our sample.}
\label{monte}
\end{figure}

\section{X-RAY EMISSION AND AGN ACTIVITY}
\label{xray}
The best independent evidence of nuclear AGN activity in the 
nuclei of our sources arises if the
source is identified as a luminous ($L_X>10^{42}$ erg/s) X-ray source.
X-ray data for a total exposure time of $\sim$1 Msec obtained with the
\textit{Chandra} observatory in the GOODS/CDFS field have been
published and made publicly available by Giacconi et al. (2002, see
also Alexander et al. 2003).
We have searched for individual X-ray emission from the 20 objects
in the present sample by cross-correlating the IRAC positions of our
targets with the positions of X-ray sources as catalogued by
Giacconi et al. (2002) and Alexander et al. (2003).

We found that three sources (objects \#6, \#11 and \#13) were
individually detected in the X-rays; for all of them the X-ray/IRAC
positional difference was smaller than 2$''$.
The three sources share similar X-ray properties: all of them have soft
X-ray spectra and have been detected only in the soft (0.5-2 keV) or full
(0.5-8 keV) X-ray band, with $\sim13-30$ counts, i.e. at the
limiting flux of the X-ray observation ($\sim 1\times10^{-16}$
erg cm$^{-2}$ s$^{-1}$). The large values observed for the X-ray to
optical flux ratio (F$_X$/F$_{opt}$) are about 10-100 times higher
than generally found for AGNs and make these objects ``extreme'' with
respect to the overall X-ray source population.
The basic X-ray properties, as drawn from Alexander et al. (2003), are
reported in Table~\ref{table_xray}.

\begin{figure*}[!ht]
\centerline{\psfig{file=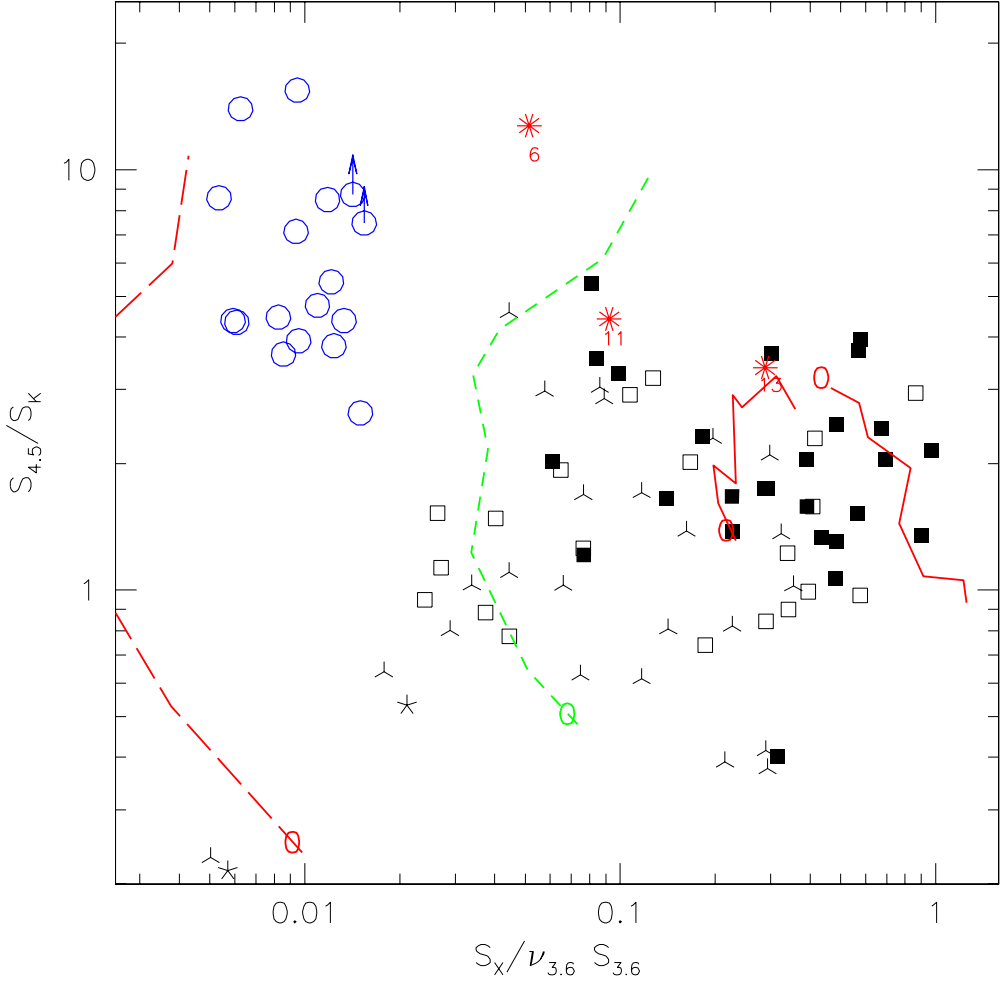,width=14cm}}
\caption{Plot of the ratio of the 3.6 $\mu$m to total X-ray fluxes
  against the 4.5 to $K$-band monochromatic fluxes. Our sample sources
  are the red asterisks (the 3 X-ray detected objects) and the open
  circles (the 17 individually X-ray undetected sources).   For the
  latter we report the X-ray to 3.6 $\mu$m flux ratio assuming for all
  17 objects the $2.3\ 10^{-17}$ erg/cm$^2$/s total average flux
  detected through the "`stacking"' analysis.
These data are compared with data on an X-ray selected sample by
  Franceschini et al. (2005), including type-1 and type-2 AGNs (filled
  and open squares) and type-2 objects dominated by the host galaxy
  emission (3-legged stars).
The lines are the predicted broadband colours as a function of $z$;
  the colour for $z = 0$ corresponds to the point marked with 0, and
  lines are drawn from here to points corresponding to z = 0.5, 1,
  1.5, 2, 2.5, 3, and 3.5. From right to left, the lines correspond to
  type-1 quasar and Seyfert-1 (the two red solid lines), Seyfert-2
  (the green short-dashed line), and Sb spiral/starburst (red
  long-dashed line).
}
\label{X-IR}
\end{figure*}

\begin{table*}[!ht]
\caption{X-ray detected sources in our sample}
\begin{tabular}{rlcccclc}
\hline
$\#$ & X-ray ID$^{a}$ & $\Delta(X-IRAC)$ & Counts$^{a}$ & Counts$^{a}$
& Counts$^{a}$ & $L_{0.5-10 keV}$$^{b}$ & best-fit $z_{\rm phot}$ \\
\hline
\hline
11-  4569 & 217 & 1.90$''$ & 17.1$^{+6.9}_{-5.7}$ & $<15.2$ & $<14.2$ &
1.1$^{+0.4}_{-0.4} \times 10^{43}$ &$4$  \\
13-  5021 & 205$^{c}$ & 0.21$''$ & 43.8$^{+9.0}_{-7.8}$ &
29.6$^{+7.2}_{-6.0}$ & $<15.5$ & 2.7$^{+0.5}_{-0.5}\times10^{43}$ & 4 \\
6- 10945 & 232$^{c}$ & 0.06$''$ & 17.4$^{+8.0}_{-6.8}$ &
13.7$^{+6.2}_{-5.1}$ & $<12.8$ & 3.6$^{+1.6}_{-1.4}\times10^{43}$ &
$8\ddag$  \\
\hline
%\enddata
\tablenotetext{a}{X-ray data from Alexander et al. (2003)}
\tablenotetext{b}{Intrinsic X-ray luminosity at the best-fit
photometric redshift, from Table 2}
\tablenotetext{c}{These sources are also present in the Giacconi et al. (2002)
catalogue, as XID \# 217 and XID \#557, respectively, with very similar
X-ray parameters}
\tablenotetext{\ddag}{
Following Zheng et al. (2004), Mainieri et al. (2005) associate the observed
X-ray emission of \#6 (XID \#557 in their papers) to a faint optical
source at a larger distance from the X-ray centroid ($\sim 1.4''$), and derive
a photometric redshift $z=1.81$.
However, the detection of a significant 2.2 $\mu$m signal in correspondance
of the IRAC centroid make object \#6 the most likely counterpart to the X-ray sources
(see Figure \ref{10945}).
}
\end{tabular}
\label{table_xray}
\end{table*}

The low number of counts prevented us to perform detailed analyses of
the X-ray spectral properties of the sources, especially about the
amount of gas absorption, given that at $z\sim4$ ($\sim8$), the 0.5-2
keV observed band corresponds to an intrinsic 2.5-10 keV band (5-20
keV band) and is thus sensitive only to column densities larger than
5$\times10^{22}$ cm$^{-2}$ (5$\times10^{23}$ cm$^{-2}$).
With a conservative approach, we derived the intrinsic X-ray
luminosities, assuming an unabsorbed power-law with $\Gamma=1.8$ at
the best-fit source redshifts, and using XSPEC (Version 11.3.1) to
translate the observed count rates into rest frame 0.5-10 keV luminosities. 
The results are reported in Table 2.
In all the cases, the X-ray luminosities are in excess of 10$^{43}$
erg s$^{-1}$, therefore suggesting that nuclear activity is fuelling
the central engine.

Seven sources with a solid X-ray detection ($>25$ counts) and
undetected at the limit of the GOODS/CDFS observations were already
reported by Koekemoer et al. (2004, hereinafter K04), who named these
objects Extreme X-ray to Optical objects (EXOs).
All of them were subsequently detected in the IRAC channels (Koekemoer
et al. 2005) with best-fit photometric solutions in the range $2<z<5$
(only one being a candidate $z>7$ AGN).
Only one of the three X-ray detected objects in our sample, \#13, is
present in Table 1 of K04. The other two escaped the K04 selection
because of the low number of X-ray counts ($<20$).
Conversely, of the remaining six objects in the K04 sample, five
escaped our selection criteria in the IRAC 3.6 $\mu$m flux (K04 source
\#4, fainter than 1.8 $\mu$Jy) or in the $K$-band flux (K04 sources
\#2,3,5,7 have $K_{\rm AB}<=23.5$ in our photometry), while K04 source
\#1 was not considered because of a blending problem in the IRAC 3.6
$\mu$m image.
The fainter $K$-band and X-ray fluxes of the X-ray detected objects in
our sample with respect to the published EXOs fit well in the narrow
$K$-band to X-ray flux correlation shown in Brusa et al. (2005), and
can be an effect of a higher redshift origin.

In order to constrain the average X-ray properties of the remaining 17
individually undetected sources, we have applied the ``stacking
technique'', following Nandra et al. (2002) and Brusa et al. (2002).
For the photometry, a circular aperture with a radius of $2''$ centered at the positions of our sources was adopted.
The counts were stacked in the standard soft, hard and full bands (0.5-2 keV, 2-7 keV, and 0.5-7 keV).
Extensive Monte Carlo simulations (up to 10,000 trials) have been
carried out by shuffling 17 random positions and using the same
photometry aperture (2 arcsec). The random positions were chosen to
lie in ``local background regions'' to reproduce the actual background
as close as possible.
The resulting distributions for the trials are shown in Fig.~\ref{monte}.

No signal has been detected in the hard band, for a total effective
exposure time of 14.6 Ms. An excess of counts above 3$\sigma$ of the
expected background level was instead clearly detected in the soft and
full bands. Assuming an unobscured $\Gamma=1.8$ power-law spectrum,
the stacked count rate in the 0.5-7 keV band (2.46e-6 cts s$^{-1}$, or
$2.3\ 10^{-17}$ erg/cm$^2$/s) corresponds to an average 0.5-10 keV
rest-frame luminosity of $\sim 1.2 \times 10^{42}$~erg~s$^{-1}$, at a
z$=4$ representative redshift.
We have verified, by splitting the analysis into various subsamples,
that this stacked signal was not due to a few objects brighter than
the average, but was uniformly spread in the sample.

We plot in Figure \ref{X-IR} the ratio of the 3.6 $\mu$m to total
X-ray fluxes against the 4.5 to $K$-band monochromatic fluxes. Our
sample sources are the 3 X-ray detected objects (red asterisks) and
the 17 individually X-ray undetected sources (blue open circles).   For
the latter we report the X-ray to 3.6 $\mu$m flux ratio assuming for
all 17 objects the $2.3\ 10^{-17}$ erg/cm$^2$/s total average flux
detected through the "`stacking"' analysis.
These data are compared with data on an X-ray selected sample by
Franceschini et al. (2005), including type-1 and type-2 AGNs (filled
and open squares) and type-2 objects dominated by the host galaxy
emission (3-legged stars).
The lines are the predicted broadband colours as a function of $z$ for
a type-1 quasar and a Seyfert-1 (the two red solid lines), a Seyfert-2
(the short-dashed line), and Sb spiral/starburst galaxy (red
long-dashed line).

Fig. \ref{X-IR} confirms that 3 of our high-z galaxies almost
certainly contain an obscured AGN, because they fall in the
AGN-dominated region, and in view of their large inferred X-ray
luminosities.
The remaining individually undetected objects fall in a region
intermediate between that of star-forming galaxies and of
AGNs. Fig. \ref{X-IR} indicates some probable AGN contributions in the
bulk of our high-$z$ galaxy sample. These results are then not
inconsistent with those of Figs. \ref{stern} and \ref{webb}.

\section{CANDIDATE MASSIVE GALAXIES AT VERY HIGH REDSHIFTS}

\subsection{ID-6: A Very Massive Candidate Galaxy at $z\sim$8?}
\label{z8}

\begin{figure*} %[!ht]
\centerline{
\psfig{file=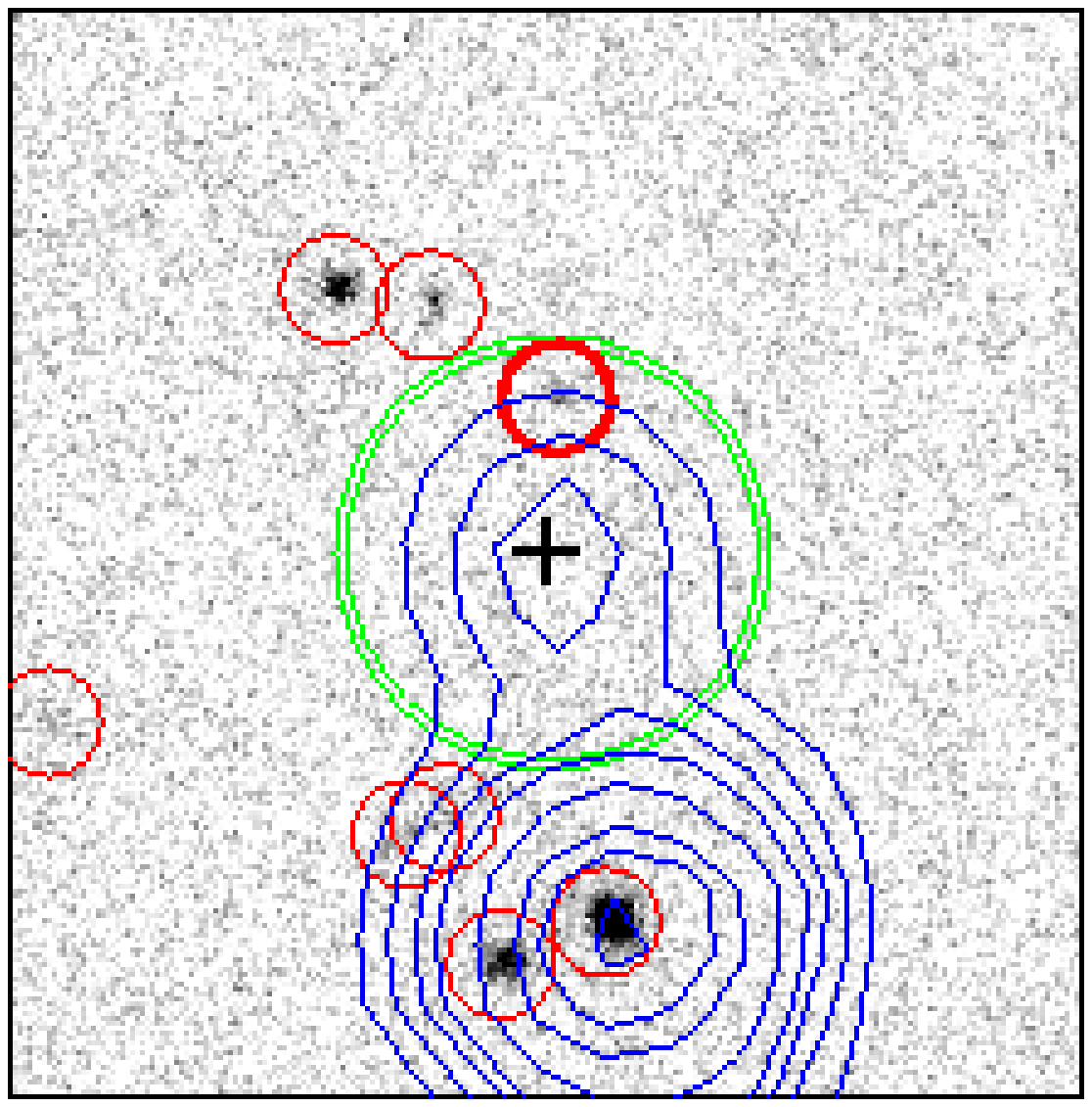,width=8cm}
\psfig{file=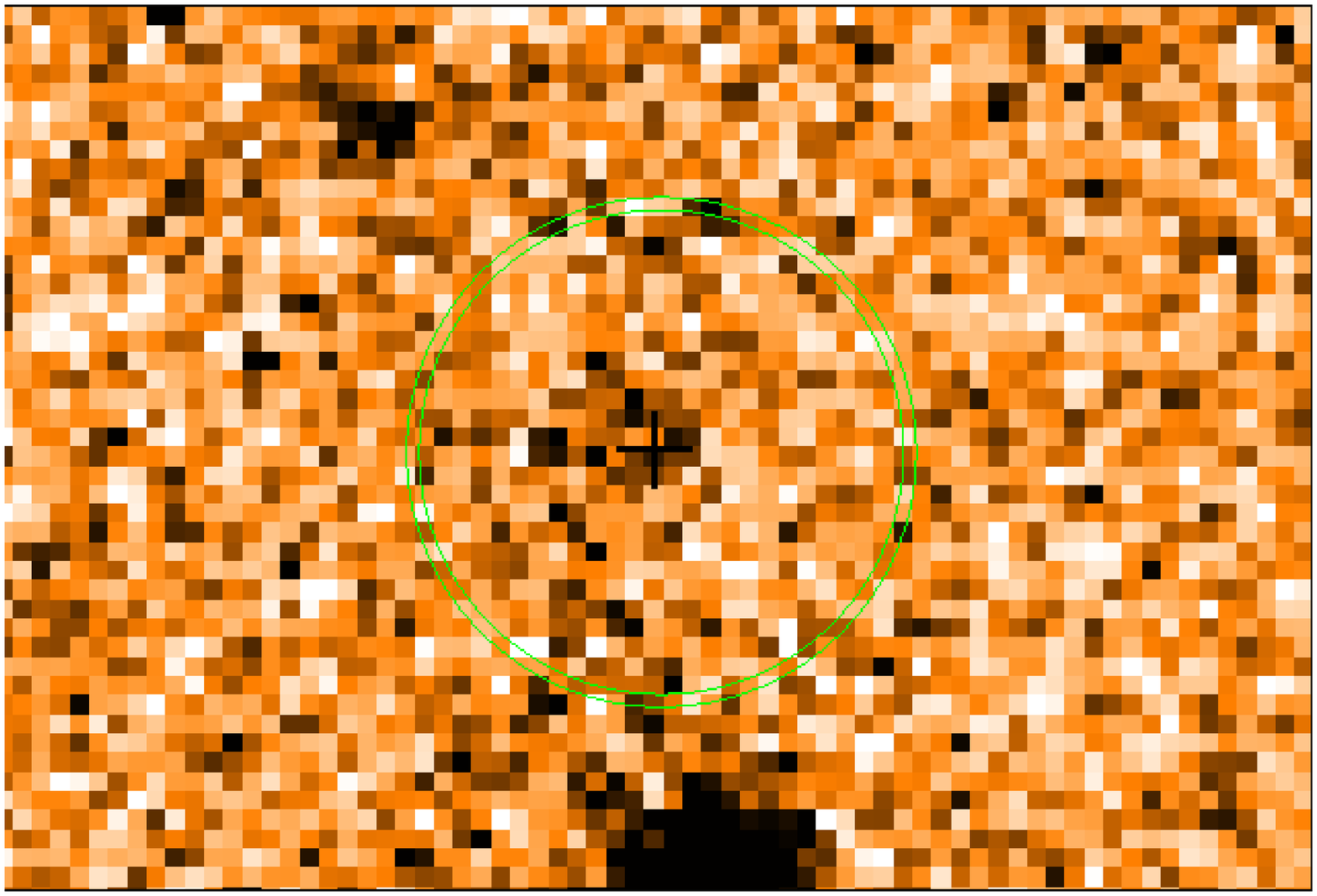,width=8cm}
}
\caption{Identification of source \#6. Left panel: the cutout shows a
10''$\times$10'' negative map of the HST $z$-band.
The black cross marks the centroid of the 3.6 $\mu$m detected source,
while the corresponding IRAC contours are shown in blue (2, 3, 4, 5,
6, 7, 10, 12, 15$\sigma$).
The green circle (2" radius) represents the position of the X-ray
source as reported by Alexander et al. (AID \#232).
The red circles (1'' diameter) indicate the optical sources detected
in the field. The thicker red source is the optical counterpart for
the X-ray source reported by Zheng et al. (2004) and Mainieri et
al. (2005). Right panel: a 10''$\times$7'' map of the ISAAC $K$-band,
with overlaid the circle indicating the position of the X-ray
source (2" radius). There is a significant 2.2 $\mu$m signal in correspondance
with the IRAC centroid, excluding the identification with other
sources in the optical map of the left panel.
}
\label{10945}
\end{figure*}

We mentioned in Section \ref{z-phot} that the SED of source \#6
formally favours a photometric redshift at $z\sim8$, with a
mass of the order of $M\sim8-10\times 10^{11} M_{\odot}$.
Given the potential relevance of this finding, we further analyse this object here.

\begin{figure*}[!ht]
\centerline{
\psfig{file=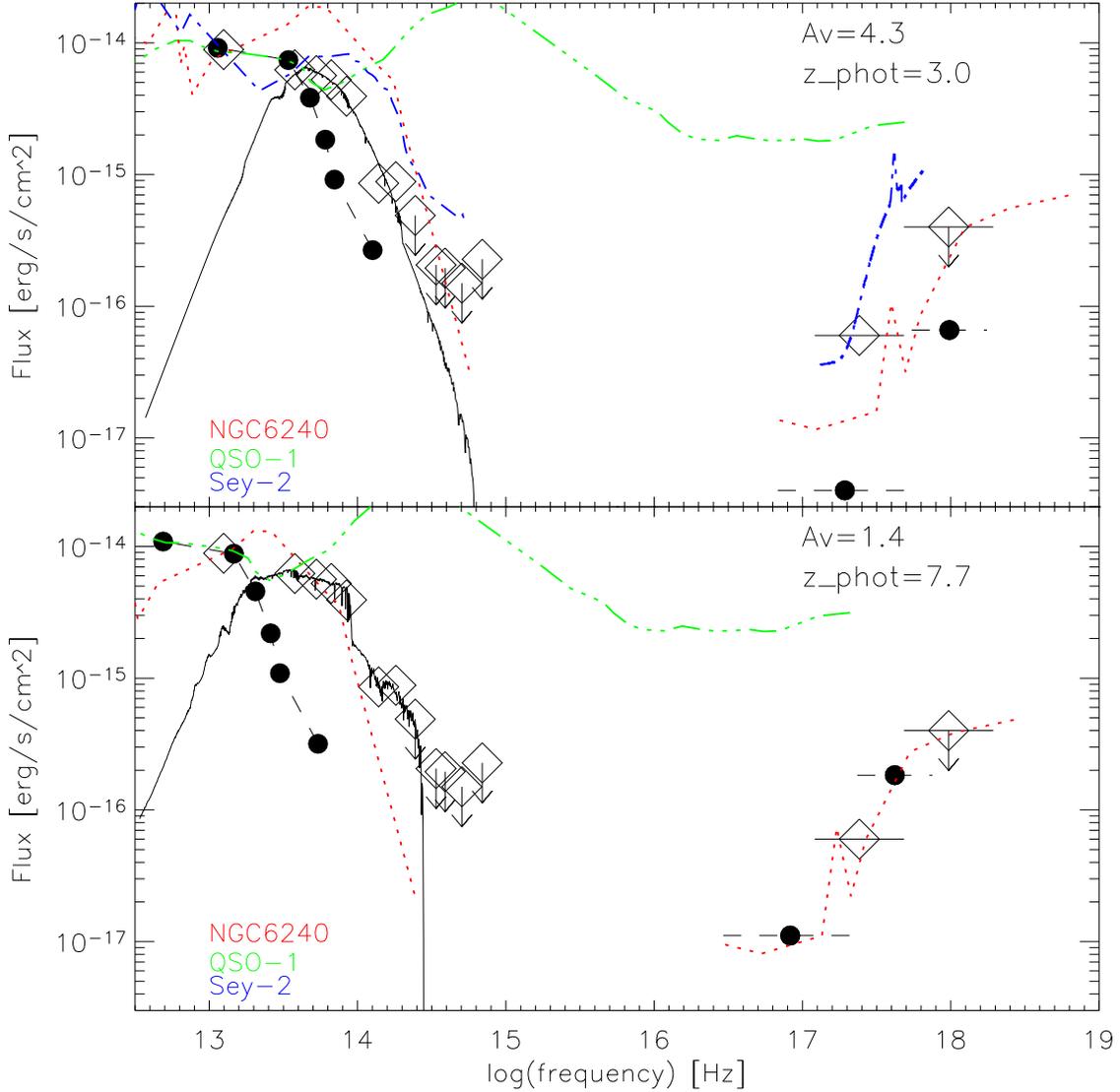,height=17cm,angle=0}
}
\caption{Two main possible solutions for object ID\#6.
In the lower panel we consider the primary solution at $z\sim7.7$, while in the
upper panel we show for comparison a secondary solution at $z\sim3.0$.
The stellar component of the source (solid black lines) in both cases
correspond to the solution already explored in Section \ref{z-phot}
and in Figures \ref{flz1}-\ref{fhz1}.
We compare the observations with various prototype
templates: a type-1 QSO (three dots-dashed green lines, Elvis et al.,
1994), the classic type-2 QSO NGC6240 (dotted red lines, Hasinger et
al. 2001), and a typical Seyfert 2 (dot-dashed blue line).
The models have been redshifted at the
corresponding photometric redshifts and normalized to fit the
observed 24 $\mu$m flux.
We have also compared the two solutions with the recently discovered
most luminous Compton-thick AGN at $z\sim2.5$ (SWIRE
J104409.95+585224.8, Polletta et al. 2006).
The SED of this object (redshifted and normalized to the 24 $\mu$m
flux) is represented by the filled black circles.
}
\label{10945sed}
\end{figure*}

Figure \ref{10945} details the multi-wavelength identification of the
source. The \textit{Chandra} X-ray object associated to this source,
in particular, has been differently identified by various authors. The left panel in the
figure shows a 10''$\times$10'' map of the HST $z$-band overplotted to
the IRAC contours shown in blue. The black cross marks the centroid of
the 3.6 $\mu$m detected source.
The (2" radius) double-lined circle represents the position of the
X-ray source as reported by Alexander et al. (AID \#232), while the
red circles (1'' diameter) indicate the optical sources detected in
the field. The thicker circle marks the optical counterpart for the
X-ray source reported by Zheng et al. (2004) and Mainieri et al. (2005).
We report in the  right panel of Fig. \ref{10945} a postage stamp of
the $K$-band image, showing that there is a faint but significant 2.2
$\mu$m source in correspondance with the IRAC centroid and excluding
the identification with other sources in the optical map (all at
$>1.8$ arcsec distances).  In addition, the X-ray and IRAC centroids
are spatially coincident and clearly indicate the presence of an IR
and X-ray source at this position which is completely absent in the
optical images.

We have then further investigated the SED properties of this
object by combining its IR and X-ray information.
The complete observed SED of the source is reported in Figure \ref{10945sed}.
In the lower panel we consider the primary solution at $z\sim7.7$, while in the
upper panel we show for comparison a secondary solution at
$z\sim3.0$. 
This latter was not formally excluded by our analysis, on the basis
of the $\chi^2$ statistics discussed in Section \ref{z-phot}:
a secondary minimum is present at this lower redshift in both
panels of Fig. \ref{chi_cont1}. We discuss also this solution in the 
present section.

In Fig. \ref{10945sed} we compare the SED data for ID\#6 with various redshifted
prototype spectral templates: a type-1 QSO from Elvis et al. (1994),
the type-2 QSO NGC6240 (Hasinger et al. 2001), and a typical Seyfert 2
galaxy, all spanning the entire X-ray-to-infrared wavelength
range. For both redshift solutions, the models
have been normalized to fit the observed 24 $\mu$m flux.
The black continuous-line spectrum from the near-IR to the UV corresponds to
the integrated stellar component solution for the galaxy already found
in Sect.\ref{z-phot} and Figs. \ref{flz1}-\ref{fhz1}.
It is clear that a purely stellar spectrum cannot explain the large
observed flux excess at 24$\mu$m. In the low-$z$ solution, this excess
can be explained as due to dust re-radiation from either a starburst
or an AGN, whereas in the high-$z$ hypothesis the only explanation would
be a dusty AGN emission.
In either case, a type-1 QSO spectrum is entirely inconsistent with the data.

This simple analysis provides interesting hints on the nature of
source \#6: in particular, if we refer to the spectral template of NGC
6240 normalized at 24 $\mu$m, the observed X-ray flux is best
explained within the higher-z solution, with a redshift of $z\sim7.7$.
The low-z, $z\sim 3.0$, fit shows instead a difficulty in explaining
the X-ray data for both the NGC 6240 and the Seyfert-2 templates,
since in both cases there is a mismatch of the 24$\mu$m to X-ray flux
ratio, and excess emission would be expected in the optical/near-IR.

\begin{figure*}[ht!]
%\centerline{
\psfig{file=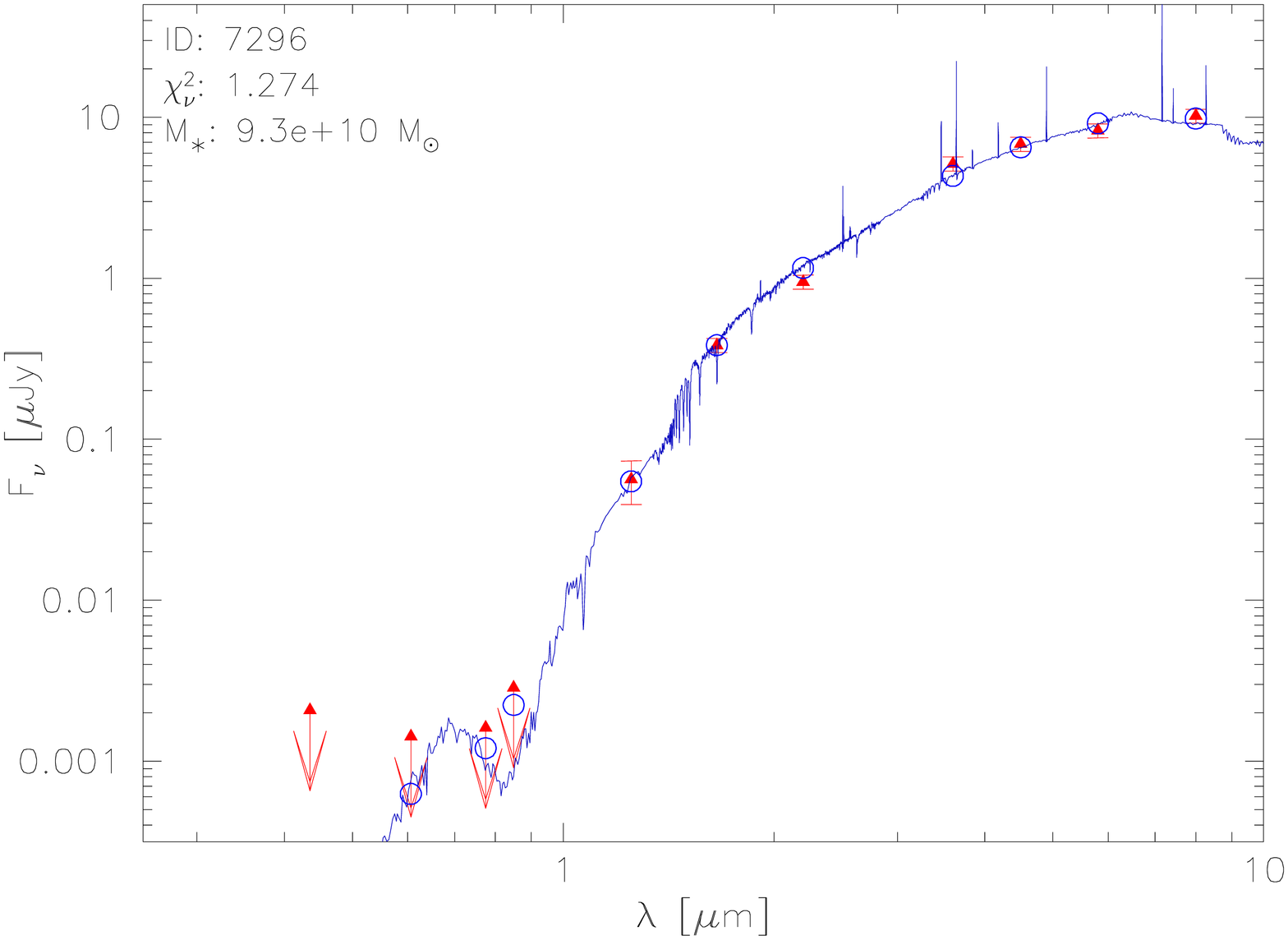,angle=270,width=14cm}
%\centerline{
\psfig{file=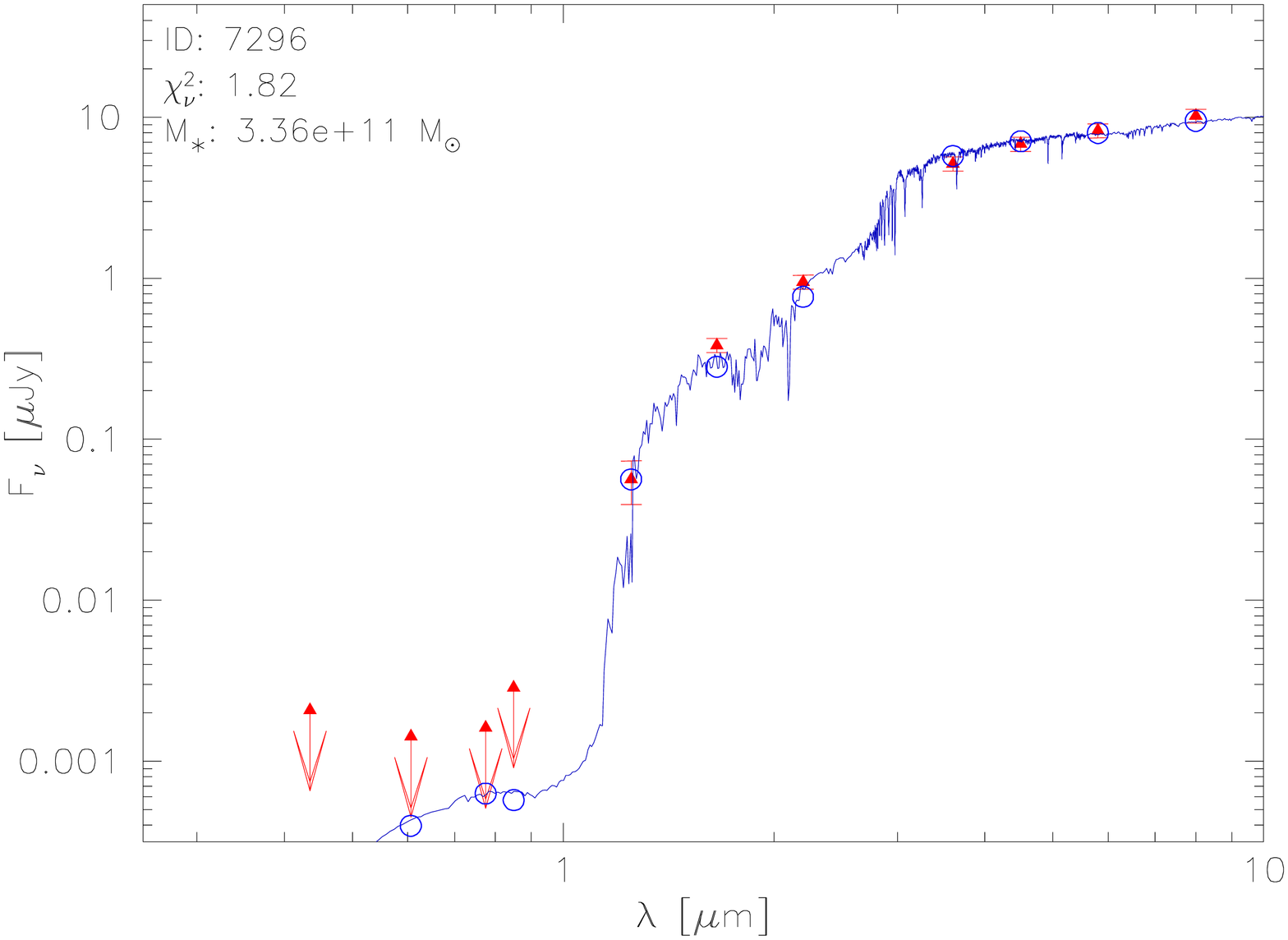,angle=270,width=14cm}
\caption{SED analysis for source \#5 (Mobasher's et al. source) based on a detailed photometric model.
The upper panel corresponds to the $z$=2.82 solution, the lower panel to the $z$=6.5 solution.
The observed photometric data are marked as red-triangles.
The $best-fit$ model is represented with the blue lines.
The convolution of the model in the various photometric bands is shown
as open blue circles. Note that in this case we used Mobasher's
photometry based on the UDF, in order to directly compare their
results with our  modellistic predictions.
}
\label{mobasher_fit1}
\end{figure*}

A very interesting comparison of the two redshift solutions is also
possible with data on the most luminous Compton-thick quasar recently
discovered at $z\sim2.5$ (SWIRE J104409.95+585224.8) by Polletta et
al. (2006).
The SED of this object (redshifted and normalized to the observed 24
$\mu$m flux) is shown as black filled circles in both panels of
Fig.\ref{10945sed}. Again, for the low-$z$ (lower panel) case, the
expected quasar X-ray flux would be too faint and undetectable in
X-rays.
When redshifted to $z\sim7.7$, instead, this quasar spectrum turns out
to fit remarkably well the far-IR and X-ray data of source \#6: the
very high redshift brings the unabsorbed hard X-ray spectrum into the
observational soft X-ray wavebands. The optical and near-IR emissions
would in any case be due to the stellar component of the host galaxy
(solid line, see Sect.\ref{z-phot}).

The analysis presented in this Sect. provides only a qualitative
support to the case for a very high-redshift solution for source \#6,
which would imply the existence of an extremely massive galaxy
($M\geq8\times10^{11} M_{\odot}$) at $z\sim 8$. Clearly, given the
potentialy extraordinary nature of this object, the lower-redshift $z\sim3$
solution apparent in Fig.8  must still be considered the more
conservative conclusion.  Unfortunately, proving this alternative 
might turn out impossible before the advent of JWST and
ALMA.   The implications of such a finding will be discussed in the
next Sections.

\subsection{ID-5: The Mobasher et al. Candidate High-z Galaxy}
\label{Mobasher}

Our object \#5 has been originally identified and analysed by Mobasher
et al. (2005, object HUDF-JD2), and subsequently discussed by Dunlop
et al. (2006) and Fontana et al. (2006), among others.   The
availability of extremely deep photometric data from UDF and near-IR
spectroscopy prompted us to perform a more detailed analysis of this source.
To this end we adopted a model by Fritz et al. (2007,in preparation;
see also Berta et al. 2005 and Poggianti et al. 2000). The model
spectrum is obtained by summing SSP spectra that are weighted by
different mass values. Each SSP is extinguished with extinction values
that are allowed to vary as a function of age (selective dust
extinction).
Dust is assumed to be distributed in a uniform screen ($R_V=3.1$).
The model computes also the equivalent widths of all relevant
interstellar emission lines.
Nine SSP spectra have been assumed with ages that were allowed to vary
from $10^6$ to $\sim 3\cdot 10^9$, that of the older stars being
chosen so as to be consistent with the age of the universe.
The best fit model was found with a $\chi^2$ analysis after
convolution with the filter response functions.
Furthermore, a constraint was used on the H$\alpha$ emission based on
the near-IR spectroscopic observations by Mobasher et al.

Making use of the photometric data by Mobasher et al. we explored both the low- and high-$z$ solutions for source \#5,
running our photometric code in the redshift ranges $1<z<3$ and
$5<z<8$.
In the low-$z$ case, we used as a further constraint the H$\alpha$ 
line flux upper limit, which we have taken to be $9\times10^{40}$ erg/s 
from Mobasher et al.

Our code found two $best-fits$ with photometric redshifts $z\sim2.82$
and $z\sim6.5$ in the two considered intervals. These values are fully
consistent with our preliminary investigations (Sect.\ref{z-phot}).
The two solutions are shown in Figure \ref{mobasher_fit1}. The upper
panel corresponds to the $z$=2.82 solution, the lower panel to that at
$z$=6.5.
The $\chi^2$ statistics sligthly favours the $z\sim2.82$ solution
($\chi^2=1.27$), but the $z\sim6.5$ scenario is equally acceptable
($\chi^2=1.82$).
We have found, in particular, that the lack of any $H\alpha$ 
emission signal may still be brought into
consistency with this low-$z$ interpretation if selective dust
extinction (i.e. extinction as a decreasing function of the age of the 
various contributing stellar populations) is taken into account.

\begin{figure}[ht!]
\centerline{
\psfig{file=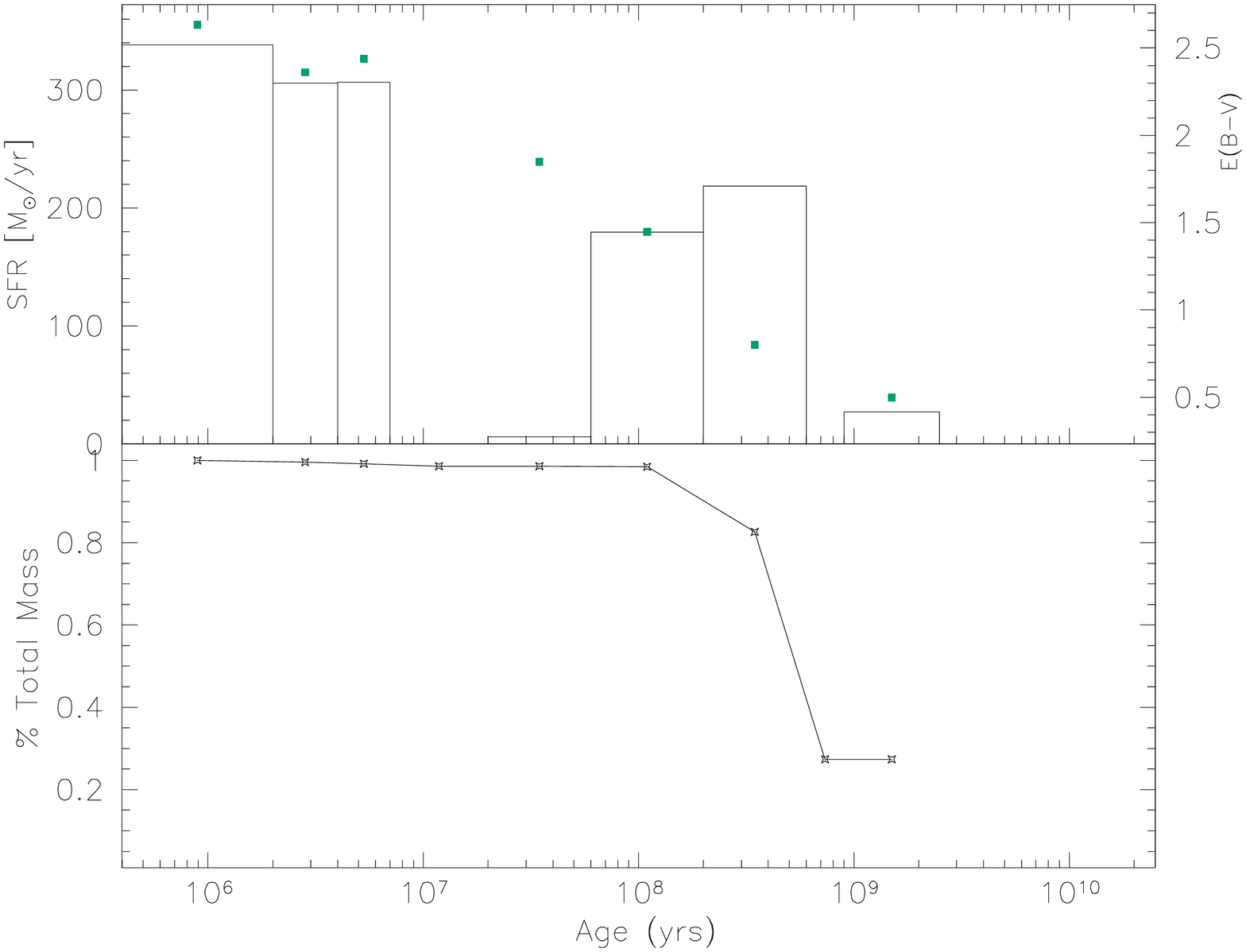,angle=270,width=9.5cm}  }
\caption{
Upper panel: Star formation history of  source \#5 for the $z=2.82$
solution. The extinction of each SSP populations is also reported
(green  points) on the right vertical axis. Lower panel: cumulative
distribution of the stellar mass assembled in the galaxy as a function
of the SSP age.
}
\label{mobasher_fit2}
\end{figure}

The $z$=6.5 fit corresponds to a single-burst solution, with an age
of 0.3 Gyr, that we already found (Sect. \ref{z-phot}). In Figure
\ref{mobasher_fit2} we show the SF history retrieved by the model for
the $z$=2.82 solution of Fig. \ref{mobasher_fit1}.   In this case the
spectrum is reproduced by SSP ranging from $2\ 10^9$ to $10^6$ yrs and
extinction as a strong function of the SSP age. The bulk of the
stellar mass would be produced between $10^9$ and $10^8$ yrs before
the observation.
In view of all these considerations, the lower redshift solution for 
galaxy \#5 appears as a more likely explanation.

\section{MAIN IMPLICATIONS}
\label{discuss}

The results of the analysis presented in this paper are relevant in the
general framework of galaxy formation and evolution, and their main
implications are hereby outlined.

\subsection{Massive Galaxies at $z\sim$4?}
\label{z4_sol}

Four sources in our sample (\#8, \#11, \#18  and \#20) show a best-fit 
photometric redshifts around $z\sim4$. Solutions at lower redshift
produce significantly worse $\chi^2$, although they are still formally
acceptable.
In addition, two more objects (\#4 and \#19) from the bimodal sample
present all the acceptable solutions at high redshift, with fits of 
comparable quality at $z\sim4$ and $z\sim6$, and very low extinction. 
We note here that only source \#4 is detected at 24$\mu$m.

All these $z\sim4$ candidates share similar physical properties with each
other. Assuming that $z\sim4$ is the correct interpretation, in most cases
the spectral best-fit solutions correspond to either the single-burst model
(see Sect. \ref{z-phot}) or to an exponentially declining SFR with
short decay time-scale $\tau=0.1$ Gyr.
The typical ages are of the order of $\sim$1 Gyr or younger with the BC03 models
($\sim$1.4 Gyr with the MA05 model).
The SEDs are then consistent with galaxies at
$z\sim4$ observed several hundred million years after a powerful burst of
star formation producing stellar masses around $10^{11}\ M_{\odot}$
with both the MA05 and BC03 models, and for a  Chabrier IMF.

At lower redshifts, numerous massive ($M>10^{11} M_{\odot}$) and old
(1-4 Gyr) early-type galaxies have been recently detected from $K$-band
selected  surveys in the range $1<z<2$ (Cimatti et al. 2004, McCarthy et
al. 2004, Glazebrook et al. 2004, Daddi et al. 2005, Saracco et
al. 2005, Kriek et al. 2006). Similar detections at slightly higher
redshift have also been recently reported ($z\sim3.7$, Brammer \& van
Dokkum, 2007)
The photometric modelling generally suggests that these passive systems
should have formed all their stars at $z_f>$3 in a short burst.

Our sources appear to be among the oldest stellar systems at
$z\sim4$, given that the estimated ages are close to the age of the
Universe at that redshift.
The post-starburst nature of our $z\sim 4$ sources and their typical
large stellar masses around $M\sim10^{11} M_{\odot}$ suggest that
they could be seen as the progenitors for the most massive spheroids
that are observed around $z\sim2$.
Moreover, as already mentioned, the
$z\sim4$ population is mostly undetected at 24 $\mu$m, making the
passively  evolving nature of these sources a plausible hypothesis.

One of this galaxies (\#11, see Sect.5 and Fig.\ref{X-IR})
has been detected in X-rays and is consistent with hosting an
obscured AGN. X-ray emissions from all the other 5 objects mentioned
in this Section have been investigated by us via a \textit{stacking}
analysis, which has detected significant average signal (at better than 3$\sigma$).

\subsection{Extremely dusty starbursts at $2<z<3$?}
\label{z2_sol}

As mentioned in Sect. \ref{z-phot}, the remaining fraction of the sources
in our sample show a strong bimodality in the photometric redshift
solutions. We also showed there that these objects are consistent with
a population  of heavily dust-enshrouded starbursts at redshift
$2<z<3$, since only few of them are undetected at
24 $\mu$m .  Based on this flux, we have computed the instrinsic
bolometric luminosity, $L_{IR}= L(8-1000 \mu m)$ (see discussion in
Sect. \ref{z-phot}.

The derived luminosities qualify most of these sources as ULIRGs
($L_{IR}>10^{12}\,L_{\odot}$). Assuming that the mid-IR emission is mostly
contributed by star-formation processes, we can translate the IR luminosity
into a SFR adopting the standard relation of Kennicutt et al. (1998).
The derived SFRs have a median value $\sim 3000 M_{\odot}/yr$, even greater
than the typical value of submillimetre selected galaxies at $z\sim2.2$
($\sim 1700 M_{\odot}/yr$, Chapman et al. 2004).
The large bolometric luminosities ($L_{IR}>10^{12} L_{\odot}$) implied
by our photometric analysis and the
instrinsic large extinction required ($A_V\sim2-4$), make our sample
very similar to the optically obscured sources identified by Houck et
al. (2005) and could represent the most extreme cases of dusty
galaxies detected until now.

It seems likely, however, that part of this IR emission may be due to an
AGN contribution, if we consider that the average X-ray signal and the
corresponding X-ray-to-3.6 $\mu$m flux ratios in Fig. \ref{X-IR}
reveal X-ray activity in excess of that expected from star-forming
galaxies.

In spite of the intrinsic uncertainties in the photometric solutions,
we have computed the contribution of these sources to the SFR density of
the Universe by assuming that their far-IR emission is mainly contributed by star formation
processes. We have made us for this of the $1/V_{max}$ estimator (see more details in Sect.
\ref{mass_density}).
Even including all objects with a photometric-$z$ solution in the redshift range
$1<z<3.5$ and detected at 24 $\mu$m (\#1, \#2, \#3, \#5, \#7, \#9, \#10, \#14, and \#16), we found that these sources missed by optical surveys contributes for only $\sim$20\% to the
global SFR density at $1<z<3.5$. We obtained a value of
$\rho_{SFR}$=0.026 $M_{\odot}/yr/Mpc^3$ (see Perez-Gonzalez et al. 2005 for a
recent compilation of the Lilly-Madau diagram of IR-selected galaxies).

\subsection{Massive Evolved Post-Starburst Galaxies at $z>6$?}
\label{z6_sol}

The fourteen sources with bimodal photometric redshift solutions
(upper panel in Table 3) have all a secondary best-fit in the redshift range $4.5<z<9$, see
Sect. \ref{z-phot} and Fig. \ref{fhz1}, and appear to have been
detected less than 1 Gyr after a powerful burst of star formation
producing a stellar mass  of the order of  $\sim 10^{11}M_{\odot}$.

Two sources in particular, (\#4 and \#19) have formally better
solutions at $z>6$ and predicted masses $M>10^{11}M_{\odot}$ from both
evolutionary synthesis models.
In addition, source \#6 has a best-fit at $z\sim7.7$.
These three sources have predicted ages for their stellar populations of
$<500$ Myr at their estimated photometric redshifts of 6.8, 6.2 and 7.7,
in the order.
Would all these sources be at $z>6$ galaxies, then these would have
formed the bulk of their massive stellar populations at very high
redshift, $z>9$, in a period of several tens million years before
entering a quiescent phase. This would have required an enormous star
formation activity, with rates of the order of
$1000-6000M_{\odot}/yr$.

Two objects (\#4 and \#6) have MIPS detections at 24 $\mu$m and one is bright
in X-rays (\# 6), which could only be interpreted at such high redshifs 
as the signature of a luminous type-2 obscured quasar within their nucleus (this
conclusion being proposed also by Mobasher et al. for source
\#5/HUDF-JD2).
Mobasher et al. also estimated the size of the dark matter halo
required to host the stellar mass of source \#5/HUDF-JD2 and found a
value of $4\times10^{12}M_{\odot}$ (if the estimated stellar mass of
the source is $6\times10^{11}M_{\odot}$ at $z\sim6.5$ and a Salpeter
IMF is considered).

Assumed such high redshifts and the associated huge energetics from
both star formation and black-hole gravitational
accretion, this would have had an important role in the reionization
of large surrouding volumes of the Universe, starting the process at
redshifts as high as $z\sim15$ (Panagia et al. 2005).
The detection of three comparably massive sources within the same
small sky region ($\sim$130 square arcmin) would support the
hypothesis that the reionization of the Universe might be dominated by
such massive galaxies.

However, two of the very high-redshift galaxy candidates have
lower-$z$ best-fit solutions of comparable quality,
which are obviously much more likely. Only for source \#6 the high-z 
($z\sim8$) fit is formally preferred to the low-z one (Sect. \ref{z8}).

At this stage, and waiting for more decisive future observations
with JWST and ALMA, we can only set a (stringent) limit to the
existence of massive evolved galaxies at very high redshifts.

\section{CONTRIBUTION TO THE COMOVING STELLAR MASS DENSITY AT HIGH Z}
\label{mass_density}

Given the significance of the massive galaxy population at
$z\sim4$ (Sect. 7.1), and even that of the $z\sim6$ candidates
(Sect. 7.3), it is worthwhile to attempt to compare their
contribution to the global comoving stellar mass density with those
already derived from other independent surveys at those redshifts. We
have computed it by using the $1/V_{max}$ estimator (see
e.g. Franceschini et al. 2006 for an application).
Our evaluation should be considered a lower limit to the
comoving mass density because our sample is not purely flux-limited
(we have excluded $a~priori$ blended sources and have applied various
colour limits, see Sect. \ref{selection}).

{\bf (A) $z\sim4.$} We have limited our analysis to the more massive
galaxies, those with $M>10^{11} M_{\odot}$. For the galaxies described in Sect. 7.1, 
we have computed for each sources the effective co-moving volume of the survey
(defined by the survey area and the redshift interval, $z_{min} < z <
z_{max}$, within which each of the $z\sim4$ candidates could
have been detected, with $z_{min}=3.7$). To determine $z_{max}$, we have
taken the best-fit SED to each candidates and redshifted it until its flux
falls below our limit ($S_{3.6}=1.8 \mu$Jy), and taken as $z_{max}$ the
minimum of the corresponding redshift and our upper boundary of the redshift
interval, z=4.7 (for most of our objects $z_{max}=4.7$, corresponding to a
survey volume $V_{max}=3.9 \times 10^5 Mpc^3$ for our adopted cosmology).
We have derived two estimates of the mass density based on the MA05 and
BC03 spectral solutions reported in Table 3.

\begin{figure}
\centerline{
\psfig{file=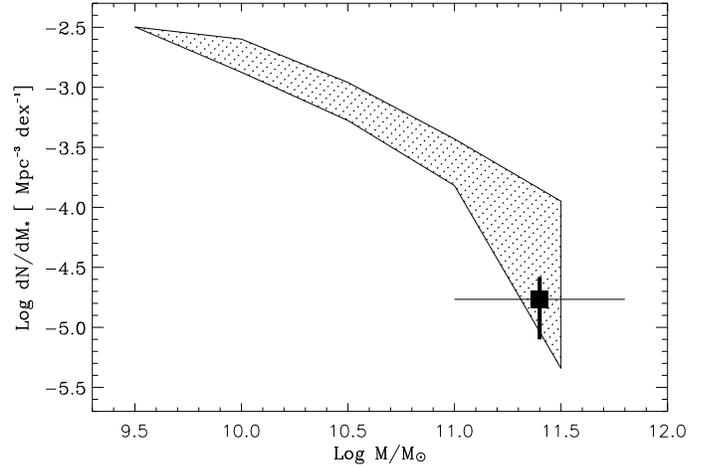,width=9.5cm}
}
\caption{The stellar mass function at $z\sim4$.
The dashed region corresponds to the data reported by Drory et
al. (2005). We have considered the mean value of the mass function in
the two redshift bins $3<z<4$ and $4<z<5$. The lower and upper
envelops of the dashed region correspond to the values derived in the
GOODS/CDFS and Fors Deep Field, respectively, by Drory et al. (2005).
The filled square marks our estimated lower limit to the mass function
at $z\sim4$ for our sample: the range spanned by the vertical arrow
corresponds to the range of values computed with MA05 (lower boundary) and BC03 (upper boundary) models.
}
\label{mf}
\end{figure}

\begin{figure}
\centerline{
\psfig{file=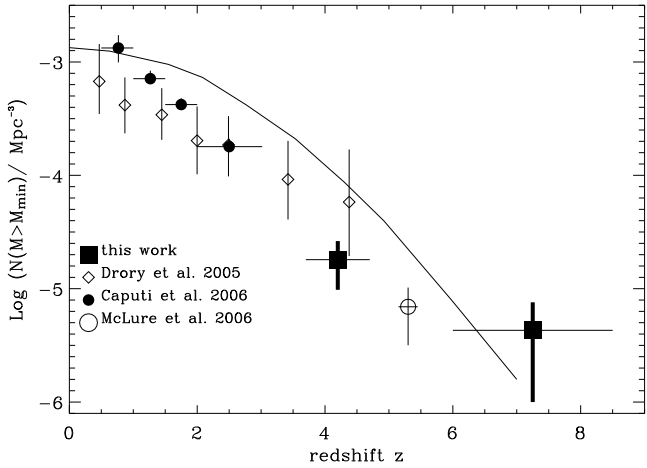,width=9.5cm}
}
\caption{Redshift evolution of the number density of galaxies with
stellar masses $M>10^{11} M_{\odot}$: our data are reported with the same symbol
as in Figure \ref{mf}, at $z=4$. The open circle at $z=5.3$ is the
estimate for LBGs with $M>10^{11} M_{\odot}$ recently derived by McLure et
al. (2006). The filled circles are the estimated number densities of
galaxies with $M>10^{11} M_{\odot}$ from the GOODS/CDFS $K$-band selected sample
of Caputi et al. (2006). We also report the compilation by Drory et
al. (2005, open diamonds) in the same mass range. The
asteriks marks the number density of {\it red and dead} galaxies with
$M>0.5\times 10^{11} M_{\odot}$ at $2<z<3$ detected by Labb\`e et al. (2005).
The solid line shows the redshift evolution of the number density of
dark matter halos with masses of  $M>2\times10^{12} M_{\odot}$,
matching the number density of galaxies with $M>10^{11} M_{\odot}$ at
$z=0$ (adapted from Figure 5 of McLure et al. 2006).
}
\label{ndens}
\end{figure}

We report in Figure \ref{mf} the contribution of our galaxies to the
stellar mass function at $z\sim4$.    The shaded region corresponds to
the data reported by Drory et al. (2005), of which we have considered
an average of their mass functions in the two redshift bins $3<z<4$
and $4<z<5$. The lower and upper envelops of the region correspond to
the mass functions derived by Drory et al. in the GOODS/CDFS and FORS
Deep Field, respectively.
The filled square marks our estimated contribution to the mass
function by our $z=4$ galaxies.
The vertical thick errorbar corresponds to our overall uncertainty, in
which the lower limit corresponds to stellar mass values computed with
MA05 (based only on 3 objects with favoured photo-$z$ at $z$=4, \#11,
\#18, and \#20), while the upper
bound is based on BC03 fits (including 3 sources with favoured
photo-$z$ at $z$=4,\#11, \#18, and \#20, and additional 7 galaxies having
bimodal photo-$z$ solutions, one of which solutions is within
$3.7<z<4.7$, \#1,\#3,\#4, \#12,\#14,\#15 and \#19, see Sect.\ref{z2_sol} above).

Figure \ref{ndens} compares our estimated number density at $z=4$ of
galaxies more massive than $M=10^{11} M_{\odot}$ with literature data
at different redshifts.
The open circle at $z=5.3$ is the estimate for LBGs with $M>10^{11}
M_{\odot}$ recently derived by McLure et al. (2006), filled circles
are from the GOODS/CDFS $K$-band selected sample of Caputi et
al. (2006), and open diamonds from Drory et al. (2005) in the same
mass range.
Altogether, our very red galaxies account for a large fraction of the
galaxy mass density at $z=4$ and are among the most massive galaxies
currently known at such redshifts.

{\bf (B) $z\sim6.$} As already discussed, our conclusions at higher
redshift are subject to major uncertainties because of the
degeneracy in the solutions for half of our sample (Sections \ref{z6_sol} and \ref{z-phot}).
However, a maximal and minimal case for the contribution to the
stellar mass density implied by our analysis may give interesting
insight.
For the galaxies described in Section 7.3 with a photometric solution
in the redshift range  $5.5<z<8.5$, we then estimated such
contribution as for the $z\sim4$ case, again, we have limited our
analysis to the more massive galaxies, those with $M>10^{11}
M_{\odot}$.
The result of this computation is shown in Figure \ref{ndens} as the
filled square at $z\sim6.7$.
The vertical thick errorbar corresponds to the overall uncertainty, in
which the lower limit corresponds to stellar mass values computed only
with source \#6 (the unique object in our sample with a clearly
favoured solution at high redshift), while the upper bound includes
six sources (\#4, \#5, \#6, \#10, \#12, \#16, and \#19).

The search for $z\sim4$ galaxies has so far been mainly performed with the
traditional \textit{Lyman dropout} colour selection technique (Steidel et al. 1999).
Lyman Break Galaxies (LBGs) are associated with starburst galaxies at
high redshifts, identified by the colours of their far ultraviolet
spectral energy distribution around the 912 $\AA$ Lyman continuum
discontinuity (Giavalisco et al. 2002).
It has been recently suggested that 60\% of the stellar mass at $z
\sim5$ is missed by the traditional drop-out selection technique
(Stark et al. 2006), which exclude high-redshift galaxies too red in
the rest-frame UV to fall under the LBG selection (McLure et al. 2006).

A fraction of our galaxies might represent a complementary sample of mature and red
galaxies which would, by definition, not be selected by the \textit{Ly
  break} or other methods making use of the optical-UV rest-frame
emission. Our results in Figs. \ref{mf} and \ref{ndens} show that the
contributions by this new, previously unaccounted, population is
definitely non-negligible if not to dominate the galaxy mass function
at the highest $z$.
These objects, all escaping detection by
published optically-selected or K-band selected samples, would almost
double the estimated value for the high-end of the galaxy mass
function (Drory et al. 2005; Fontana et al. 2006) at that redshift.

\begin{figure}
\centerline{
\psfig{file=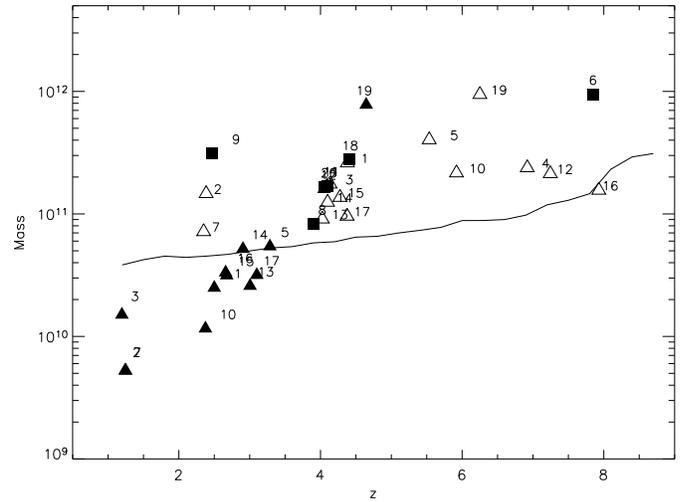,width=9.5cm} }
\caption{Plot of the stellar mass versus photometric redshift,
  summarizing the results of our selection for very high-$z$ galaxies
  based on selection at IRAC 3.6 $\mu$m in the 130 \textit{arcmin$^2$}
  GOODS CDFS field, a K-band flux fainter than $K=23.5$ AB mag, and
  non-detection in any GOODS optical bands.
For 11 galaxies shown in filled squares we show the unique photo-z
  solution, for the remaining 9 galaxies we both report as filled
  triangle the lower-$z$ fit and with the open triangle the higher-$z$
  one.  The continuous line corresponds to the mass of a 1 Gyr-old SSP
  detectable at the survey limit.
The selection procedure clearly tends to detect massive evolved galaxies around $z\sim4$.
}
\label{summary}
\end{figure}

Would this reassessment of the galaxy mass function still be consistent with standard
$\Lambda$CDM model expectations at such high redshifts?
Following McLure et al. (2006) and Dunlop et al. (2006), we report in Fig.
\ref{ndens} the redshift evolution of the number density of
dark matter halos with masses of  $M>2\times10^{12} M_{\odot}$ (soli line),
matching the number density of galaxies with $M>10^{11} M_{\odot}$ at
$z=0$ (adapted from figure 5 of McLure et al. 2006).
The proposed model corresponds to a ratio between halo and stellar
mass of $\sim20$ (see discussion in McLure et al. 2006).
Looking at Fig. \ref{ndens} we can see that the number density of
our massive sources at $z\sim4$ is a factor $\sim$2-8 (with MA05 and
BC03, respectively) lower than the predicted number of dark matter halos.
Even if our data is only a lower limit, this implies that our
estimated number density of massive galaxies at $z\sim4$ is still
consistent with the current predictions of $\Lambda$CDM
models for the hierarchical formation of cosmic structures.

\section{DISCUSSION AND CONCLUSIONS}
\label{conclusion}

Recent observational progress has revealed the existence of massive
structures (galaxies and galaxy aggregates) at high redshift much more
frequently than originally supposed.
Indeed, the problem for any attempts to predict
the origin of galaxies from first principles is dealing with the very
complex physical processes involving baryons (e.g. Somerville et al. 2001,
Somerville et al. 2004).
Observations have also shown an unexpected trend for the most massive
part of the galaxy mass function to be put in place first, and lower
mass galaxies to keep forming stars at lower redshifts (Cowie et
al. 1996; Franceschini et al. 1998, 2006; Bundy et al. 2005).

At the present stage, however, the timing for and the physical
processes accompanying the emergence of massive galaxies with cosmic
time, which would be so informative for our understanding of galaxy
formation, are essentially unknown. A problem here is that the most
efficient tool to identify very high-redshift galaxies, the
\textit{Ly-dropout} technique, is not sensitive to galaxy mass but
rather to UV flux.

The advent of sensitive imagers in the near-IR atmospheric JHK bands and particularly of the \textit{Spitzer}
IRAC space facility have started to provide new powerful selection
tools more sensitive to the host stellar mass. Several reports have
been recently appeared about searches for massive galaxies at $z>4$
and the evolution of the stellar mass function (Stark \& Ellis 2006; Yan
et al. 2006; McLure et al. 2006; Grazian et al. 2006; Fontana et
al. 2006).  Most of these have in any case exploited the
\textit{Ly-dropout} approach and pushed it to the reddest optical bands
for the highest-$z$ characterization and used IRAC data to constrain
the stellar mass.

Dunlop et al. (2006; see also Mobasher et al. 2005) followed a
complementary method of carrying out an extensive SED-fitting analysis
on large flux-limited samples without preconceived assumptions about
the rest-frame spectrum of the candidates, and by using as reference a
K-band selection with $K_S<23.5$. They conclude in favour of a lack of
evidence for very massive galaxies to be in place at $z>4$.

We follow a similar approach here, but extend it to a galaxy selection
based on the most sensitive \textit{Spitzer} 3.6 $\mu$m IRAC images in
the 130 \textit{arcmin$^2$} GOODS CDFS, a $K$-selection complementary to
that of Dunlop et al. ($K>23.5$ AB mag), and non-detection in any
GOODS optical bands.
By these means we were aiming at detecting the most
massive and highest redshift galaxies in the field to the 3.6 $\mu$m
limiting flux, and keeping equally sensitive to dusty star-forming and
massive evolved galaxies at high $z$.

Our results in terms of the stellar mass and redshift for such extreme 3.6
$\mu$m selected galaxies are summarized in Figure \ref{summary}, where
we plot the stellar mass versus photometric reshift estimated from SED
fitting. Data for the 6 galaxies with favoured single photo-z solution in our discussion, are
shown as filled squares, while for the remaining 9 galaxies results of
both the lower-$z$ fits (filled triangles) and the higher-$z$ ones
(open triangles) are reported.

Given the uncertainties in the photometric redshift solutions, our
results are consistent with previous reports, and point towards
very few of the galaxies in the field being found at $z>6$, with only
one such extreme case (galaxy \#6) being formally indicated (though a lower-$z$
fit is still acceptable). Two
other, similarly massive, galaxies are consistent with $z>6$ solution
but do not require it.
Remarkably, and similarly to what found by Dunlop et al., two of the three
candidates have solid detections in the \textit{Spitzer} MIPS 24
$\mu$m band. However, differently from Dunlop et al., we do not
consider this as necessarily an argument in favour of a lower-$z$
case, because the presence of an obscured AGN could easily explain
it. Indeed, we have found some evidence for (optically) hidden AGNs in
the majority of our sample of very red high-redshift galaxies from the
ultra-deep \textit{Chandra} X-ray data (Fig. \ref{X-IR}), which adds
to the frequently observed mid-IR 24 $\mu$m excess.
We concluded from this that there might be room for substantial
contribution to re-ionization to happen in relation with
star-formation and AGN activity in massive galaxies.

One major result of our analysis is about the potential existence of a
candidate population of massive galaxies detected around redshift 4. The
majority  of our sample galaxies (14 out of 20) have a photo-$z$ solution at
$3.7<z<4.7$, and 4 of them have best-fit solutions in this
redshift interval (not formally unique, however).
Hence, in spite of the small numbers, a galaxy population undetected in 
the optical and extremely faint in the $K$-band appears to possibly 
dominate the massive end of the galaxy
mass function at $z=4$. These objects, all escaping detection by
published optically-selected or K-band selected samples, would almost
double the estimated value for the high-end of the galaxy mass
function (Drory et al. 2005; Fontana et al. 2006) at that redshift.

Several of these evolved $z\sim4$ galaxies (none of the 4 with robust
$z=4$ solution, but 4 of the 7 with "`secondary"' solution at such $z$)
display strong excess emission at 24 $\mu$m. This result is similar to
that reported by Dunlop et al. (2006) for their selected very high-$z$
population. Due to the large redshifts, this would correspond to
rest-frame emission at 4-5 $\mu$m, hence would be difficult to explain
purely as dust reprocessing by star-forming regions. Again this result
requires rather common AGN activity in these high-$z$ evolved
galaxies. A support to this interpretation comes from the deep X-ray
data, revealing 2 galaxies with \textit{bona-fide}
$z=4$ photometric-$z$ to have clear AGN-like emissions (with
$L_X>10^{43}\ erg/s$), and the remaining objects also showing excess
X-ray flux (Fig. \ref{X-IR}), although at a lower level.
Some evidence for probable AGN contributions at 8 and 24 $\mu$m was
also found directly in the colour-colour plots of Figs. \ref{stern}
and \ref{webb}.

This widespread association of very high-$z$ galaxies with trace
obscured AGN activity might bring an interesting confirmation of the
emergent view (e.g. De Lucia et al. 2005; Bower et al. 2006; Granato et
al. 2004) that AGN feedback could have systematically influenced the
shaping of the galaxy mass function during the epoch of galaxy
formation.

\vspace{0.75cm} \par\noindent
{\bf ACKNOWLEDGMENTS} \par
\noindent
This work is based on observations made with the {\it Spitzer Space Telescope},
which is operated by the Jet Propulsion Laboratory, California Institute of
Technology under NASA contract 1407.
ESO/GOODS observations have been carried out using the Very Large
Telescope at the ESO Paranal Observatory under Program ID(s): LP168.A-0485 .
The NASA/ESA $Hubble~Space~Telescope$ is operated by the Association
of Universities for Research in Astronomy (AURA), Inc., under NASA
contract NAS5-26555.
We thank the referee for his/her detailed comments and suggestions
that improved the quality of our work.
We warmly thank Claudia Maraston from providing us with her evolutionary model
predictions in electronic form and Alvio Renzini, Stefano Berta and Vincenzo
Mainieri for useful discussion.

{}

%\appendix
%\label{jac}

%\section{SED analysis with a detailed spectro-photometric model}

\landscape
\begin{table}
  \caption{Summary of best-fit parameters from $Hyperz$:  photometric
    redshift, extinction ($A_V$), age, mass and IR luminosity.
The upper panel of the Table refers to those sources showing a bimodality
in the photometric redshift solutions (14 objects).
For each parameter we present the lower redshift primary solution (sol. I)
and the secondary solution (sol. II) at higher redshift.
The lower panel refers to sources with a single favoured solution.
For each physical parameter we report a range of values: the value
reported on the left side of the interval corresponds to the prediction
of the BC03 model, while the value reported on the right side of the
interval corresponds to the prediction of the MA05 model.
  }
\centerline{
\begin{tabular}{|cccccccccccc|}
%\begin{tabular}{|c||cc||cc||cc||cc||cc||c|}
\hline
~&{\bf BIMODAL}& {\bf solutions}&~&~&~&~&~&~&~&~&~\\
\hline
  ID   &  $z$-phot &$z$-phot & AV &AV & $age [Gyr]$ & $age [Gyr]$ & Mass [$log(M_{\sun})$] & Mass [$log(M_{\sun})$]  &$\chi^2_{\nu}$  &$\chi^2_{\nu}$ & $L_{bol}$ [$log(L_{\sun})$]$^a$  \\
~      &sol I &solII  &sol I &solII  &sol I &solII &sol I &solII  &sol I &solII  &sol I \\
\hline
%\tiny
1  & 2.70-2.30 & 4.40-4.36 & 5.80-5.80 & 1.20-0.40& 0.004-0.004& 1.015-1.434& 10.403-10.160 & 11.421-11.181 & 4.410-5.266 &  2.160-1.985 &13.80\\
\hline
2  & 1.28-1.20 & 2.36-2.41 & 8.10-8.10 & 2.00-1.80& 0.006-0.006& 3.500-2.600&  9.812- 9.400 & 11.106-10.995 & 1.832-2.033 &  1.441-1.492 &12.95 \\
\hline
3  & 1.00-1.39 & 4.87-3.40 & 7.50-7.80 & 0.00-0.00& 0.004-0.002& 1.015-1.700&  9.871-10.253 & 11.275-10.982 & 2.961-2.886 &  5.485-9.025 &12.67 \\
\hline
4  & 3.98-4.20 & 6.88-6.96 & 1.50-0.90 & 0.80-0.00& 1.015-1.434& 0.509-0.509& 11.205-11.027 & 11.468-10.055 &15.213-3.255 & 12.373-2.721 &-- \\
\hline
5  & 3.24-3.33 & 6.41-4.66 & 4.70-4.70 & 0.20-0.00& 0.006-0.004& 0.509-1.434& 10.782-10.456 & 11.611-11.369 & 4.356-4.905 &  1.285-2.834 &14.04 \\
\hline
7 & 1.29-1.21 & 2.35-2.35 & 8.30-8.00 & 2.40-1.60& 1.434-0.006& 3.500-2.600& 10.096-9.114 & 10.822-10.658 & 1.145-1.003 &  1.201-1.089 &10.75 \\
\hline
10 & 2.75-2.00 & 7.08-4.76 & 5.80-5.80 & 0.00-0.00& 0.004-0.006& 0.719-1.434& 10.252-9.646 & 11.391-11.048 & 1.700-2.476 &  0.992-3.027 &13.64 \\
\hline
12 & 4.00-4.10 & 7.50-6.99 & 1.40-0.40 & 3.40-3.40& 1.015-1.015& 0.002-0.002& 11.278-10.902 & 11.289-11.137 & 0.286-0.382 &  1.458-2.112 &14.03 \\
\hline
13 & 3.00-3.01 & 4.37-3.68 & 4.50-4.90 & 0.60-0.00& 0.004-0.002& 1.015-1.700& 10.159-10.435 & 10.925-10.761 & 4.296-4.513 &  2.674-2.631 &-- \\
\hline
14 & 3.02-2.79 & 4.13-4.07 & 5.50-5.50 & 0.60-0.10& 0.002-0.001& 1.015-1.434& 10.501-10.697 & 11.088-10.872 & 1.437-1.401 &  0.262-0.371 &14.16 \\
\hline
15 & 2.65-2.70 & 4.52-4.03 & 1.90-4.00 & 0.30-0.00& 2.000-0.006& 1.015-1.434& 10.970-9.794 & 11.141-10.898 & 2.702-1.959 &  2.380-3.990 &13.38\\
\hline
16 & 2.69-2.63 & 7.36-8.50 & 2.00-3.00 & 1.90-2.90& 0.509-0.017& 0.509-0.002& 10.792-10.021 & 11.001-11.155 & 4.609-3.895 &  3.980-5.454 &11.61 \\
\hline
17 & 3.20-3.00 & 4.75-4.00 & 4.20-4.70 & 0.00-3.90& 0.004-0.002& 1.015-0.002& 10.252-10521 & 11.078-10.654 & 4.102-3.516 &  2.348-6.538 &13.67 \\
\hline
19 & 4.61-4.68 & 6.27-6.23 & 2.40-1.30 & 2.40-0.80& 1.015-1.015& 0.181-0.509& 11.918-11.636 & 12.038-11.681 & 4.559-4.496 &  3.549-3.603 &15.46 \\
\hline
\hline
\hline
\hline
~&{\bf FAVOURED SINGLE}& {\bf solutions}&~&~&~&~&~&~&~&~&~\\
\hline
% UNIQUE z\\
  ID   &  $z$-phot & AV & $age [Gyr]$ & Mass [$log(M_{\sun})$] & $\chi^2_{\nu}$  \\
\hline
6 & 7.79-7.90 &  1.40-0.20&  0.181- 0.360 &12.022-11.700&  1.075-0.625 \\
\hline
8  & 3.82-3.99 &  1.30-0.50&  0.509- 0.719 &10.982-10.620&  7.269-7.756 \\
\hline
9  & 2.42-2.52 &  4.10-4.10&  2.300- 2.600 &11.453-11.336&  0.343-0.343 \\
\hline
11 & 4.06-4.14 &  2.10-1.50&  0.360- 0.360 &11.292-11.001&  2.388-2.582 \\
\hline
18 & 4.84-3.98 &  0.00-0.00&  1.015- 1.434 &11.359-11.295&  2.803-5.393 \\
\hline
20 & 4.29-3.83 &  0.40-0.00&  1.015- 1.434 &11.190-11.021&  2.355-2.725 \\
\hline
\end{tabular}
\tablenotetext{a}{The instrinsic infrared luminosity of each source,
$L_{IR}= L(8-1000 \mu m)$ is estimated by scaling the known $L_{ir}$
of the best template fit to the measured MIPS 24 $\mu$m flux density
after redshifting the template SED to its measured $z$.}
}
\label{tab_highz}
\end{table}
\endlandscape

\end{document}